\documentclass[a4paper,12pt]{article}

\usepackage{epsfig}
\usepackage{a4}
\usepackage{latexsym, amssymb} 

\newcommand{\lapprox}{%
\mathrel{%
\setbox0=\hbox{$<$}
\raise0.6ex\copy0\kern-\wd0
\lower0.65ex\hbox{$\sim$}
}}
\newcommand{\gapprox}{%
\mathrel{%
\setbox0=\hbox{$>$}
\raise0.6ex\copy0\kern-\wd0
\lower0.65ex\hbox{$\sim$}
}}

\newcommand{\ba}{\begin{array}}
\newcommand{\ea}{\end{array}}
\newcommand{\bd}{\begin{displaymath}}
\newcommand{\ed}{\end{displaymath}}
\newcommand{\be}{\begin{equation}}
\newcommand{\ee}{\end{equation}}
\newcommand{\bea}{\begin{eqnarray}}
\newcommand{\eea}{\end{eqnarray}}

\def\fb{\, {\rm fb}}
\def\pb{\, {\rm pb}}
\def\met{E_T \hspace*{-1.1em}/\hspace*{0.5em}}
\def\Rp{R_p \hspace*{-0.8em}/\hspace*{0.3em}}
\def\tev{\, \, {\rm TeV}}
\def\gev{\, \, {\rm GeV}}

\newcommand{\PYTHIA}{PYTHIA}

\def\order#1{{\cal O}\left(#1\right)}
\def\gtap{\raisebox{-.4ex}{\rlap{$\sim$}} \raisebox{.4ex}{$>$}}

\def\neu {{\tilde {\chi_1}}^0}
\def\cha {{\tilde {\chi_1}}^{\pm}}

\def\mneu {m_{\neu}}
\def\mcha {m_{\cha}}

\def\mgluino {m_{\tilde{g}}}

\def\thefootnote{\fnsymbol{footnote}}

\begin{document}

\begin{titlepage}

\begin{flushright}
{\small 
September 20, 2010 \\ 
RECAPP-HRI-2010-009}
\end{flushright}

\vspace*{0.2cm}
\begin{center}
{\large {\bf Discrimination of low missing energy look-alikes at the LHC}}\\[2cm]  
Kirtiman Ghosh\footnote{kirtiman@hri.res.in},
Satyanarayan Mukhopadhyay\footnote{satya@hri.res.in}, 
Biswarup Mukhopadhyaya\footnote{biswarup@hri.res.in}\\[0.5cm]  

{\it Regional Centre for Accelerator-based Particle Physics \\
Harish-Chandra Research Institute \\
Chhatnag Road, Jhusi \\ 
Allahabad - 211 019, India}\\[2cm]
\end{center}

\begin{abstract}
The problem of discriminating possible scenarios of TeV scale new
physics with large missing energy signature at the Large Hadron
Collider (LHC) has received some attention in the recent past. We
consider the complementary, and yet unexplored, case of theories
predicting much softer missing energy spectra. As there is enough
scope for such models to fake each other by having similar final
states at the LHC, we have outlined a systematic method based on a
combination of different kinematic features which can be used to
distinguish among different possibilities. These features often trace
back to the underlying mass spectrum and the spins of the new
particles present in these models. As examples of \textquotedblleft
low missing energy look-alikes\textquotedblright, we consider
Supersymmetry with R-parity violation, Universal Extra Dimensions with
both KK-parity conserved and KK-parity violated and the Littlest Higgs
model with T-parity violated by the Wess-Zumino-Witten anomaly
term. Through detailed Monte Carlo analysis of the four and higher
lepton final states predicted by these models, we show that the models
in their minimal forms may be distinguished at the LHC, while
non-minimal variations can always leave scope for further
confusion. We find that, for strongly interacting new particle
mass-scale $\sim 600 \gev$ ($1 \tev$), the simplest versions of the
different theories can be discriminated at the LHC running at
$\sqrt{s}=14 \tev$ within an integrated luminosity of $5$ ($30$)
$\fb^{-1}$.

\end{abstract}

\pagestyle{plain}

\end{titlepage}


\setcounter{page}{1}

\renewcommand{\thefootnote}{\arabic{footnote}}
\setcounter{footnote}{0}

\section{Introduction}
\label{sec:intro}

The Large Hadron Collider (LHC) marks the beginning of an era where
physics at the TeV scale can be probed at an unprecedented level.  One
important goal of such investigations is to see whether the standard
model (SM) of elementary particles is embedded within a set of new
laws which make their presence felt at the TeV scale.  Several
proposals of such new physics (NP) have been put forward, with
motivations ranging from the naturalness problem of the Higgs mass to
solving the dark matter puzzle.

Quite a few of such models systematically predict a host of new
particles occurring in correspondence with those present in the SM.
In addition, the need to accommodate an invisible, weakly interacting
particle qualifying as a dark matter candidate often invites the
imposition of a $Z_2$ symmetry on the theory, which renders the
lightest of the new particles stable. This leads to the prediction of
large missing transverse energy (MET) at the LHC, due to the cascades
of new particles ending up in the massive stable particle that eludes
the detectors. Such MET (together with energetic jets, leptons etc.)
goes a long way in making such new physics signals conspicuous. Even
then, however, one has to worry about distinguishing among different
theoretical scenarios, once some excess over SM backgrounds is
noticed. Thus one has the task of using the LHC data to differentiate
among models like supersymmetry (SUSY), universal extra dimensions
(UED) and littlest Higgs with T-parity (LHT), all of which are
relevant at the TeV scale. With the SM particles supplemented with
new, more massive ones having the same gauge quantum numbers (with
only spins differing in the case of SUSY) in all cases, their signals
are largely similar. The consequent problem of finding out the model
behind a given set of signatures is often dubbed as the LHC inverse
problem.  The name was coined when it was shown first in the context
of SUSY~\cite{NAH} that different choices of parameters within SUSY
lead to quantitatively similar LHC signals. The efforts towards
distinction were subsequently extended to the aforementioned different
scenarios with large missing energy signature at the
LHC~\cite{look_alike,Datta_Kane,LHT_vs_SUSY_UED}.

Though the scenarios predicting MET signals are attractive from the
viewpoint of explaining the dark matter content of the universe, and
they also satisfy the electroweak precision constraints while keeping
the new particle spectrum \textquoteleft natural\textquoteright, the
discrete symmetry ensuring the stability of the weakly interacting
massive particles is almost always introduced in an {\it ad hoc}
manner.  For example, it is well-known that the superpotential of the
minimal SUSY standard model (MSSM) can include terms which violate the
conventionally imposed R-parity, defined as $R~=~(-)^{(3B+L+2S)}$. If
SUSY exists, the violation of R-parity cannot therefore be ruled
out. Similarly, boundary terms in UED can violate the Kaluza-Klein
parity usually held sacrosanct, and T-parity in LHT can be broken by
the so-called Wess-Zumino-Witten anomaly term. While one is denied the
simplest way of having a dark matter candidate when the $Z_2$ symmetry
is broken, these are perfectly viable scenarios phenomenologically,
perhaps with some alternative dark matter candidate(s). In SUSY, for
example, the axino or the gravitino can serve this purpose even if
R-parity is broken. And, most importantly, the different scenarios
with broken $Z_2$ are as amenable to confusion as their
$Z_2$-preserving counterparts, as far as signals at the LHC are
concerned. The mostly sought-after final states (such as jets +
dileptons) are expected from all of these scenarios, various
kinematical features are of similar appearance, and in none of the
cases does one have the MET tag for ready recognition, in clear
contrast to scenarios with unbroken $Z_2$.

Side by side with the problem of large missing energy look-alike
models, the disentanglement of `low missing energy look-alikes' is
thus an equally challenging issue, on which not much work has been
done yet. Some criteria for distinction among this class of
look-alikes are developed in this paper. Specifically, we consider
four possible scenarios of NP with low missing energy signature:

\begin{enumerate}
 \item Supersymmetry with R-parity violation (SUSY-RPV)
\item Minimal universal extra dimensions (mUED) with KK-parity
  conserved (UED-KKC)
\item Minimal universal extra dimensions with KK-parity violated
  (UED-KKV)
\item Littlest Higgs model with T-parity violated by the 
   Wess-Zumino-Witten anomaly term (LHT-TPV).
\end{enumerate}

It should be noted that UED-KKC is also included in this study. This
is because, as will be seen in the following sections, the peculiar
features of the mUED spectrum (namely, a large degree of degeneracy)
often leads to the lightest stable particle carrying very low
transverse momentum.  In addition, one often also has nearly back-to-back
emission of the invisible particle pair. Consequently, the MET
spectrum is rather soft over a large region of the parameter space,
and the signals can be of similar nature as those of $Z_2$-violating
scenarios. Hence we would like to emphasize that mUED with KK-parity
conserved is a scenario which can be easily distinguished by its much
softer MET spectrum from other models predicting a massive stable
particle (like SUSY with R-parity), but might actually be confused
with the other $Z_2$-violating scenarios.

It is well-known that signals containing leptons have relatively less
SM backgrounds compared to events with fully hadronic final states.
We therefore focus on possible leptonic channels in the various models
under consideration. One has to note, however, that it is difficult to
devise model-independent cuts such that the SM backgrounds are reduced
while keeping a significant fraction of the signal events intact for
all the models. For example, strong cuts on the transverse momenta of
the leptons cannot be applied in case of mUED as the leptons there are
very soft in general, owing to the almost degenerate spectrum of the
Kaluza-Klein excitations. This makes it difficult to reduce the SM
backgrounds in the single lepton and opposite-sign dilepton channels
(which have rather large irreducible backgrounds from $W+$~jets and
$t\bar{t}+$~jets respectively). Although same-sign dilepton and
trilepton channels have relatively lower rates within the SM, they are
not very suitable for the purpose of distinguishing between the above
models. The main reason for this is that many otherwise conspicuous
invariant mass peaks cannot be reconstructed in these channels. These
invariant mass peaks, however, are very helpful in classifying the
models. In addition, since one is now looking at situations where the
NP signals are {\it not} accompanied by large MET, rising above the SM
backgrounds is a relatively harder task which is accomplished better
with a larger multiplicity of leptons.  Keeping this mind, here we
have tried to develop a procedure of model discrimination, depending
on four-and higher-lepton signals as far as possible.

The four-lepton channel is viable in all the four models mentioned
above, and such events can be used to extract out several qualitative
differences among the models, including the presence or absence of
mass peaks. Therefore, our study largely focuses on the four-lepton
channel. Furthermore, since most cascades start with the production of
strongly interacting heavier particles in these new theories, as we
can be seen from the appendix, generically we can obtain
at least two hadronic jets in the signal. Apart from the four-lepton
channel, we also use the presence (or absence) of signals with even
higher lepton multiplicity as a discriminating feature. It is of
course true that these methods of discrimination can often be applied
only after sufficient luminosity has been accumulated. In this sense,
our study differs from that of \cite{look_alike}, where the inverse
problem was considered not only for models with a very large MET
signal, but also at a modest luminosity of $200 \pb^{-1}$ only.

The paper is organized as follows. The choice of relevant parameters in the various models considered and our
general strategy and methods adopted in event generation at the LHC
are summarised in section~\ref{signal-bkg}. We use this strategy to
show some predictions in section~\ref{why-alike} to convince the
reader that the different scenarios indeed fake each other
considerably at the LHC.  Section~\ref{4l-detail} is devoted to a
detailed study of kinematics of four-lepton final states, whereby a
significant set of distinction criteria are established. In
section~\ref{sec:5l}, we discuss the usefulness of the channel with
five or more leptons. Some scenarios over and above those covered here
in details are qualitatively commented upon in
section~\ref{sec:others}. We summarise and conclude in
section~\ref{summary}. Finally, as a useful reference for the reader, we outline the main features of the
low-missing energy look-alike models and the possible cascades through which four and higher-lepton signals can arise in those contexts, in an Appendix.

\section{Multilepton final states: signal and background processes}
\label{signal-bkg}
\subsection{Event generation and event selection criteria}

Before we establish that the models mentioned in the introduction (and described in the Appendix)
all qualify as look-alikes with low MET, and suggest strategies for
their discrimination through multilepton channels, we need to
standardise our computation of event rates in these channels. With
this in view, we outline here the methodology adopted in our
simulation, and the cuts imposed for reducing the SM backgrounds. The
predicted rates of events for some benchmark points, after the cuts
are employed, are presented at the end of this section.

As the production cross-section for strongly interacting particles is
high at the LHC, we start with the production of these heavier
`partners' (having different spins for SUSY-RPV, and same spins in the
remaining cases) of quarks and gluons at the initial parton level
$2\rightarrow2$ scattering. Though in principle there are
contributions to the multilepton final states from electroweak
processes as well, they are subleading, and are left out in our
estimate. In this sense, our estimates of the total cross-sections in
various channels, are in fact lower bounds. Also, the observations
made by us subsequently on final state kinematics are unaltered upon
the inclusion of these subleading effects.

For our simulation of signal processes, in case of SUSY-RPV, we have
used PYTHIA 6.421~\cite{PYTHIA} for simulating both the production of
strongly interacting particles and their subsequent decays. Since the
values of the L-violating couplings that we have taken are very small,
they do not affect the renormalisation group running of mass
parameters from high to low scale~\cite{Allanach}. Therefore, The SUSY
spectrum is generated with SuSpect 2.41~\cite{SUSPECT}. For UED, while
the production is once again simulated with the help of PYTHIA, both
KK-parity conserving and KK-parity violating decay branching fractions
are calculated in CalcHEP 2.5~\cite{CalcHEP} using the model file written
by~\cite{UED_Calchep} and then passed on to PYTHIA via the SUSY/BSM
Les Houches Accord (SLHA) (v1.13)~\cite{SLHA_SUSY_BSM}. For UED-KKV,
we have implemented the relevant KK-parity violating decay modes in
CalcHEP. Finally, for LHT with T-violation the initial parton level
hard-scattering matrix elements were calculated and the events
generated with the help of CalcHEP.  These events, along with the
relevant masses, quantum numbers and decay branching fractions were
passed on PYTHIA with the help of the SLHA interface for their
subsequent analysis. In all cases, showering, hadronization, initial
and final state radiations from QED and QCD and multiple interactions
are fully included while using PYTHIA.

To simulate the dominant SM backgrounds, events for the processes $ZZ$
and $t \bar{t}$ were generated with PYTHIA. Backgrounds from  $t \bar{t}Z$
have been simulated with the generator ALPGEN~\cite{ALPGEN}, and 
the unweighted event samples have been passed onto PYTHIA for
the subsequent analysis.

We have used the leading order CTEQ6L1~\cite{CTEQ} parton distribution
functions.  Specifically, for PYTHIA, the Les Houches Accord Parton
Density Function (LHAPDF)~\cite{LHAPDF} interface has been used. The
QCD factorization and renormalization scales are in general kept fixed
at the sum of masses of the particles which are produced in the
initial parton level hard scattering process. If we decrease the QCD
scales by a factor of two, the cross-section can increase by about
30\%.

While the signal primarily used in our analysis is $4l+nj+\met$
($n\geq2$), we have also considered events with five or more number of
leptons in section~\ref{sec:5l}. No restriction was made initially on
the signs and flavours of the leptons. As we shall see subsequently,
the `total charge of leptons' can be used as a useful discriminator at a
later stage of the analysis.

The different ways in which multilepton signals can arise in the
various models under consideration have been described in the appendix.
The principal backgrounds to the $4l+nj+\met$
channel ($n\geq2$) are $ZZ/\gamma^{*}$, $t \bar{t}$ and
$t\bar{t}Z/\gamma^{*}$. Although in case of $ZZ$ (where the two
associated jets can come from ISR and FSR), there is no real source of
MET if we demand a $4l$ signal (i.e., both the $Z$'s have to decay
leptonically), jet energy mismeasurement can give rise to some amount
of fake MET.

The following basic selection cuts were
applied for both the signal and the background~\cite{TDR,Mellado}:

\noindent
{\bf Lepton selection:}
\begin{itemize}
 \item $p_T>10 \gev$ and $|\eta_{\ell}|<$ 2.5, where $p_T$ is
   the transverse momentum and $\eta_{\ell}$ is the pseudorapidity 
   of the lepton (electron or muon). 
\item {\bf Lepton-lepton separation:}  ${\Delta R}_{\ell\ell} \geq $ 0.2,
  where $\Delta R = \sqrt {(\Delta \eta)^2 + (\Delta \phi)^2}$ is the
  separation in the pseudorapidity--azimuthal angle plane.
\item {\bf Lepton-jet separation:} $\Delta R_{\ell j} \geq 0.4$ for all jets
  with $E_T >$ 20 GeV.  
\item The total energy deposit from all {\it hadronic activity} within a cone
  of $\Delta R \leq 0.2$ around the lepton axis should be $\leq$ 10 GeV. 
\end{itemize}

\noindent
{\bf Jet selection:}

\begin{itemize}
\item 
Jets are formed with the help of {\tt PYCELL}, the inbuilt cluster routine in
\PYTHIA.  The minimum $E_{T}$ of a jet is taken to be $20\gev$, and
we also require  $|\eta_j|<$ 2.5. 
\end{itemize}

We have approximated the detector resolution effects by smearing the
energies (transverse momenta) with Gaussian
functions~\cite{TDR,Mellado}. The different contributions to the
resolution error have been added in quadrature.

\begin{itemize}
\item {\bf Electron energy resolution:}
\be
\frac{\sigma(E)}{E} = \frac{a}{\sqrt{E}} \oplus b \oplus
                      \frac{c}{E}, 
\ee
where
\be
(a, b, c) = \left\{ \begin{array}{lcl}
                   (0.030\gev^{1/2}, \, 0.005, \, 0.2\gev), & \hspace{1em} &
                                                   |\eta| < 1.5, \\ 
                    (0.055\gev^{1/2}, \, 0.005, \, 0.6\gev), & \hspace{1em} &
						   1.5 < |\eta| < 2.5. 
		    \end{array}
            \right.
\ee
\item {\bf Muon $p_T$ resolution:}
\be
\frac{\sigma(p_T)}{p_T} = \left\{ \begin{array}{lcl}
                       a , & \hspace{1em} &  p_{T} < 100\gev, \\
                       \displaystyle
                         a + b \, \log \frac{p_T}{100\gev} , & & 
			 p_{T}>100\gev, 
		    \end{array}
            \right.
\ee
with
\be
(a, b) = \left\{ \begin{array}{lcl}
                       (0.008, \, 0.037), & \hspace{1em} & |\eta| < 1.5,  \\
                       (0.020, \, 0.050), & \hspace{1em} & 1.5 < |\eta| <
                                                           2.5. \\  
		    \end{array}
            \right.
\ee
\item {\bf Jet energy resolution:}
\be
\frac{\sigma(E_T)}{E_T} = \frac{a}{\sqrt{E_T}}, 
\ee
with 
$ a= 0.5\gev^{1/2}$, the default value used in {\tt PYCELL}.
\end{itemize}

Under realistic conditions, one would of course have to deal with
aspects of misidentification of leptons and jet energy
mismeasurement. 

Note that, in addition to the basic cuts discussed above and the
further cuts on the lepton $p_T$'s described in the next sub-section,
we also have used a cut on the invariant mass of the opposite sign
(OS) lepton pairs formed out of the four-leptons. We have demanded
that $M_{l^+l^-}>10 \gev$ for all the OS lepton pairs. This cut helps
us in reducing backgrounds coming from $\gamma^{*}$ produced in
association with a Z-boson or top quark pairs. {\em We shall
  collectively refer to the basic isolation cuts and this cut on
  dilepton invariant masses as {\bf Cut-1}}.

\subsection{Numerical results for four-lepton events}
\label{num-results}
Two benchmark points have been chosen for each of the look-alike
models and it is seen that the four and higher lepton signals can rise
well above SM backgrounds, thus forming the basis of further kinematic
analyses. In order to show that our analysis is independent of the
mass-scale of new physics involved, we have chosen two benchmark
points with different mass spectra. Also, we should emphasize here that in the subsequent analysis we shall be using kinematic features of the final states predicted by the different models which are largely independent of the choice of parameters. This is precisely why we take two different benchmark points and demonstrate that the essential qualitative distinctions between the models do not change as we go from one point to another. {\em Thus our conclusions reflect distinction among the various low-MET models as a whole, and do not pertain to specific benchmark points.} The relevant parameters of these models and their values at the benchmark points are described below. For details about the models, we refer the reader to the Appendix.

For SUSY-RPV, we have worked in the minimal supergravity (mSUGRA)
framework. This is done just with an economy of free parameters in
view, and it does not affect our general conclusions. In this
framework, the MSSM mass spectrum at the weak scale is determined by
five free parameters. Among them the universal scalar ($m_0$) and
gaugino masses ($m_{1/2}$) and the universal soft-breaking trilinear
scalar interaction ($A_0$) are specified as boundary conditions at a
high scale (in this case the scale of gauge coupling unification),
while the ratio of the vacuum expectation values of the two Higgs
doublets (tan$\beta$) and the sign of the Higgsino mass parameter
(sgn($\mu$)) as defined in eqn.~\ref{eqn:mssm-sup} are specified at
the electroweak scale. In our analysis the electroweak scale has been
fixed at $\sqrt{m_{{\tilde t}_1}m_{{\tilde t}_2}}$, where ${\tilde
  t}_1$ and ${\tilde t}_2$ are the two mass eigenstates of the top
squarks respectively.

In case of the minimal universal extra dimension (mUED) model with conserved Kalutza-Klein (KK) parity (UED-KKC), the essential parameters determining the mass spectrum are the radius of compactification ($R$) and the ultra-violet cut-off scale of the theory ($\Lambda$). Although in mUED with KK-parity violated (UED-KKV) we have an additional parameter $h$, for small values of this parameter, $R$ and $\Lambda$ once again primarily determine the mass spectrum.

In the Littlest Higgs model with T-parity violation (LHT-TPV), the new T-odd quark and gauge boson masses are determined by two parameters: $f$ and $\kappa_q$ (see Appendix for their precise definitions). Apart from that, an additional parameter $\kappa_l$ appears in the T-odd lepton sector, which we have fixed to be equal to $1.0$ throughout our study.

The choice of parameters for the different models are as
follows:\\
\begin{enumerate}
 \item {\bf Point A:}
\begin{itemize}
 \item {\em SUSY-RPV}: $m_0=100 \gev$, $m_{1/2}=250 \gev$,
   tan$\beta=10$, $A_0=-100$ and sgn($\mu$)$>0$. The RPV coupling
   taken is $\lambda_{122}=10^{-3}$. With these choices, we find the
   sparticle masses relevant to our study as (all in GeV) $\mgluino=
   606$, $\mneu=97$, $\mcha=180$, $m_{\tilde{d_L}}=568$,
   $m_{\tilde{t}}=399$, $m_{\tilde{e_L}}=202$, $m_{\tilde
     {{\nu_e}_L}}=186$.
\item {\em LHT-TPV}: $f=1500 \gev$, $k_q=0.285$. With these choices,
  we find the T-odd particle masses relevant to our study as (all in
  GeV) $u_H=603$, $d_H=605$, $A_H= 230$, $W_H=Z_H=977$. 

\item {\em UED-KKC}: $R^{-1}=475$ GeV and $\Lambda=20R^{-1}$. With these
  choices of UED-KKC parameters, the masses of relevant KK-particles
  are given by (all in GeV), $m_{g_1}=609$, $m_{Q_1} =568$,
  $m_{Z_1}=509$, $m_{W_1}=509$, $m_{l_1}=489$ and $m_{\gamma_1}=476$.
  
\item {\em UED-KKV}: For UED-KKV, the values of $R^{-1}$ and $\Lambda$ are
  chosen to be same as in the case of UED-KKC. The value of the
  KK-parity violating parameter $h$ is set to 0.001. 
\end{itemize}

\item {\bf Point B:}
\begin{itemize}
 \item {\em SUSY-RPV}: $m_0=100 \gev$, $m_{1/2}=435 \gev$,
   tan$\beta=5$, $A_0=0$ and sgn($\mu$)$>0$. The RPV coupling
   taken is $\lambda_{122}=10^{-3}$. With these choices, we find the
   sparticle masses relevant to our study as (all in GeV) $\mgluino=1009
   $, $\mneu=176$, $\mcha=331$, $m_{\tilde{d_L}}=927$,
   $m_{\tilde{t}}=695$, $m_{\tilde{e_L}}=309$, $m_{\tilde
     {{\nu_e}_L}}=300$.
\item {\em LHT-TPV}: $f=2375 \gev$, $k_q=0.285$. With these choices,
  we find the T-odd particle masses relevant to our study as (all in
  GeV) $u_H=956$, $d_H=957$, $A_H=368 $, $W_H=Z_H=1549$. 

\item {\em UED-KKC}: $R^{-1}=800$ GeV and $\Lambda=20R^{-1}$. With these
  choices of UED-KKC parameters, the masses of relevant KK-particles
  are given by (all in GeV), $m_{g_1}=1025$, $m_{Q_1} =956$,
  $m_{Z_1}=851$, $m_{W_1}=851$, $m_{l_1}=824$ and $m_{\gamma_1}=800$.
  
\item {\em UED-KKV}: For UED-KKV, the values of $R^{-1}$ and $\Lambda$ are
  chosen to be same as in the case of UED-KKC. The value of the
  KK-parity violating parameter $h$ is set to 0.001.
\end{itemize}

\end{enumerate}

For LHT-TPV, UED-KKC and UED-KKV the Higgs mass has been fixed at $120
\gev$. The top quark mass has been taken as $175\gev$ in our study. In
LHT-TPV, the value of $\kappa_l$ has been kept fixed at $1$. The
choices of $f$ and $\kappa_q$ have been made in order to match the
mass scale of the strongly interacting particles in LHT with those of
the other models (these particles are predominantly produced in the
initial hard scattering at the LHC). We should note here that this
could have also been achieved with other different choices of these
parameters, and we have remarked about their implication in
section~\ref{subsec:invmass-2l}.

\begin{table}[htb]
\begin{center}
\begin{tabular}{||c|c|c|c|c||}
\hline \hline
\multicolumn{5}{||c||}{Point A}\\\hline\hline
$p_T^{l}$ Cuts& SUSY-RPV & LHT-TPV & UED-KKC & UED-KKV \\ 
(GeV)  & (fb)&(fb) &(fb) &(fb) \\
\hline
(10,10,10,10)    & 9450.91     &   21.09 &  163.36  & 14008.93 \\
\hline 
(20,10,10,10)    & 9447.87     &   21.09 &  129.29  & 13990.97 \\
\hline
{\bf (20,20,10,10)}    & {\bf 9354.09}     &  {\bf 20.96} & {\bf  75.13}  & {\bf 13819.24} \\
\hline
(20,20,20,10)    & 8486.90     &   19.66 &   30.73  & 11781.96\\
\hline
(20,20,20,20)    & 5013.02     &   12.90 &   7.16   &  7697.5 \\

\hline \hline 
\multicolumn{5}{||c||}{Point B}\\\hline\hline
(10,10,10,10)    & 756.00     &   1.71 &  26.24  & 872.94 \\
\hline 
(20,10,10,10)    & 756.00     &   1.71 &  25.83  & 872.74 \\
\hline
{\bf (20,20,10,10)}   &  {\bf 755.22}     &   {\bf 1.71} &  {\bf 22.81}  & {\bf 868.90}\\
\hline
(20,20,20,10)    & 736.43     &   1.66 &   14.73  & 804.96\\
\hline
(20,20,20,20)    & 555.70     &   1.22 &   4.77  &  520.27 \\

\hline \hline 
\multicolumn{5}{||c||}{SM Backgrounds}\\\hline\hline
                 &$ZZ/\gamma^{*}$  & $t \bar{t}$ & $t \bar{t} Z/\gamma^{*}$ & Total\\
\hline
(10,10,10,10)    & 2.18 & 0.51   & 0.87 & 3.56 \\
\hline 
(20,10,10,10)    & 2.18 & 0.51  & 0.87  & 3.56 \\
\hline
{\bf (20,20,10,10)}   &  {\bf2.17 }     &   {\bf 0.43 } &  {\bf 0.87}  & {\bf 3.47}\\
\hline
(20,20,20,10)    & 2.06     &  0.17 & 0.79  &3.02 \\
\hline
(20,20,20,20)    & 1.42     &  0.09 & 0.55  & 2.06 \\

\hline \hline 

\end{tabular}
\caption{Cut-flow table showing the change in $4l+\ge 2j+\met$
  cross-section as we gradually increase the cuts on the lepton
  $p_T$. The $p_T$ cuts shown within the brackets are
  ($p_{T}^{l_{1}}$, $p_{T}^{l_{2}}$, $p_{T}^{l_{3}}$, $p_{T}^{l_{4}}$)
  where $l_1$, $l_2$, $l_3$ and $l_4$ are the four leptons ordered
  according to their $p_T$'s. The signal (point A and point B) and the
  SM background cross-sections quoted are for $\sqrt{s}=14 \tev$. For
  all our subsequent analysis of four-lepton events, we have used the
  (20,20,10,10) cut. {\em We shall refer to this cut, applied in
    addition to Cut-1, as {\bf Cut-2}.}}
\label{tab:cut-flow}
\end{center}
\end{table}

Table~\ref{tab:cut-flow} shows the signal and SM background
cross-sections for the $4l+\geq 2j+\met$ channel at LHC running with
$\sqrt{s}=14 \tev$ after Cut-1 and additional cuts on lepton
$p_T$'s. The signal cross-sections are shown for the two above
different choices of parameters in each model. On the whole, it is
evident from the table that our event selection criteria assure us of
enough background-free events in each case. The backgrounds look
somewhat challenging for LHT-TPV for benchmark point B. However,
one still can achieve a significance (defined as $S/\sqrt{B}$, $S$ and
$B$ being respectively the number of signal and background events) 
of $5 \sigma$ with an integrated luminosity of 30 $\fb^{-1}$.
For point A, $5 \fb^{-1}$ is likely to be sufficient for
raising the signal way above the backgrounds, and attempting 
discrimination among various theoretical scenarios via kinematical
analysis.

We also show the sensitivity of the signals of the different scenarios
to gradually tightened $p_T$ cuts on the leptons. Clearly, the close
degeneracy of the spectrum in UED-KKC makes the leptons softer. Thus
the signal events fall with stronger cuts imposed on the relatively
softer leptons.  Although such cuts applied during the offline
analysis can be useful in discriminating the UED-KKC scenario from the
rest, one has to use this yardstick with caution. This is because the
situation can change with different values of the UED parameters
$R^{-1}$ and $\Lambda$ (as can be seen from point B). While they also
change the KK excitation masses together with changing the
mass-splitting, the estimate of masses from data for UED is not
reliable, for reasons to be discussed in the next section.  We
therefore suggest some kinematical criteria that bypass this caveat.

\section{On what ground are the models look-alikes?}
\label{why-alike}
In addition to the fact that all the suggested models have
  appreciable cross-section in the four lepton channel (and also in
  various other channels that we are not considering for
  model-discrimination for reasons described in the introduction), in
  order to ascertain that these models are indeed look-alikes, we
  first show that certain kinematic distributions, like MET or
  effective mass have very similar features in these scenarios of new
  physics.  To begin with, we note that with similar masses of the
  strongly interacting new particles (squarks and gluinos for
  SUSY-RPV, $q_1$ and $g_1$ for the two variants of UED, and $q_H$ for
  LHT-TPV), the MET distributions look similar.

The missing transverse energy in an event is given by 
\be \label{missing_E_T}
\met =\sqrt{\left(\sum p_x \right)^2+\left(\sum p_y\right)^2}. 
\ee
Here the sum goes over all the isolated leptons, the jets, as well as
the `unclustered' energy deposits.

\begin{figure}[h!]
\begin{center}
\centerline{
\epsfig{file=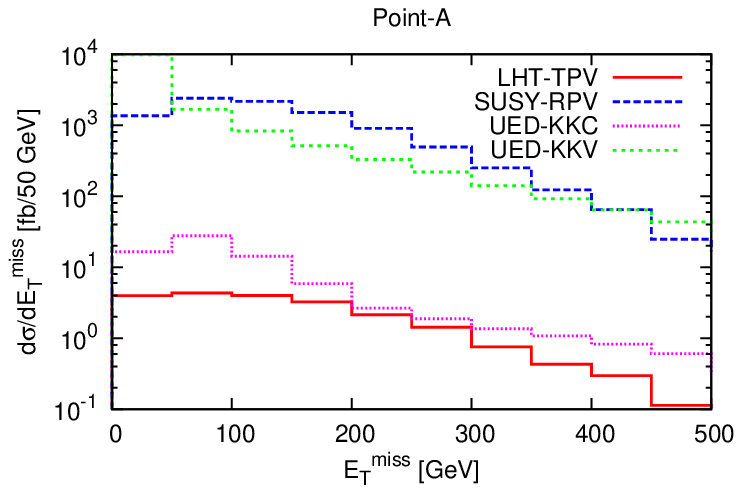,width=8.5cm,height=7.0cm,angle=-0}
\epsfig{file=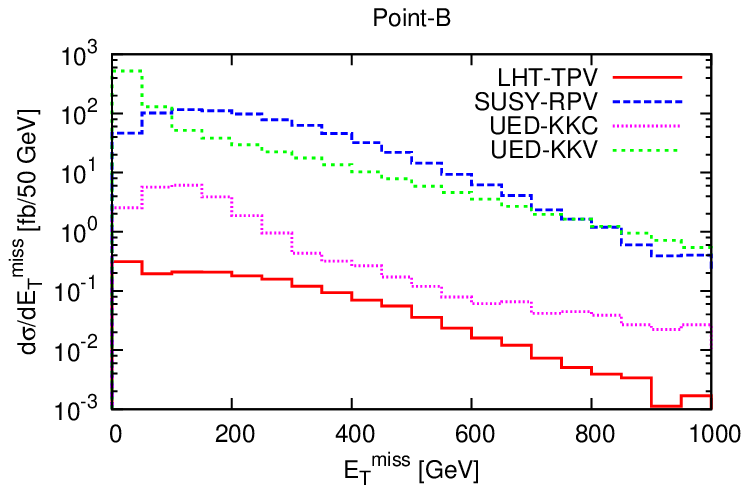,width=8.5cm,height=7.0cm,angle=-0}}

\caption{Missing $E_T$ distribution in various models for point A
  (left figure) and point B (right figure), at $\sqrt{s}=14 \tev$
  after Cut-2.}
\label{fig:MET}
\end{center}
\end{figure}

The MET distributions for the four scenarios under study are shown in
Figure~\ref{fig:MET}. The similarity of the MET distributions is
obvious in all the four cases, including UED-KKC. For UED-KKC, as
mentioned before, the reason behind this is the close degeneracy of
the spectra. As a result, in this model the two $\gamma_1$'s produced
at the end of the cascades on the two sides have little transverse
momentum, and quite often they are also back-to-back in the transverse
plane. Consequently, the net MET is considerably reduced, in spite of
the $\gamma_1$ being a massive particle. Thus UED-KKC is as much of a
look-alike with SUSY-RPV and LHT-TPV as UED-KKV.

The confusion among the look-alikes from MET distributions is expected
to be more for similar values of the masses of the strongly
interacting new particles. Since the direct measurement of mass at LHC
is not easy, a measured quantity that often bears the stamp of these
masses is the so-called effective mass. It is defined as the scalar
sum of the transverse momenta of the isolated leptons and jets and the
missing transverse energy,
\be \label{M_eff}
 M_{eff}=\sum p_{T}^{jets} +\sum p_{T}^{leptons} + \met,  
\ee 
%
\begin{figure}[h]
\begin{center}
\centerline{\epsfig{file=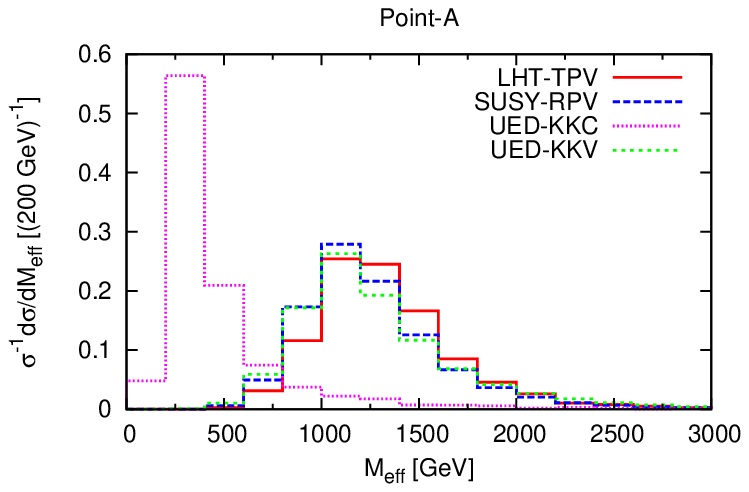,width=8.5cm,height=7.0cm,angle=-0}
\epsfig{file=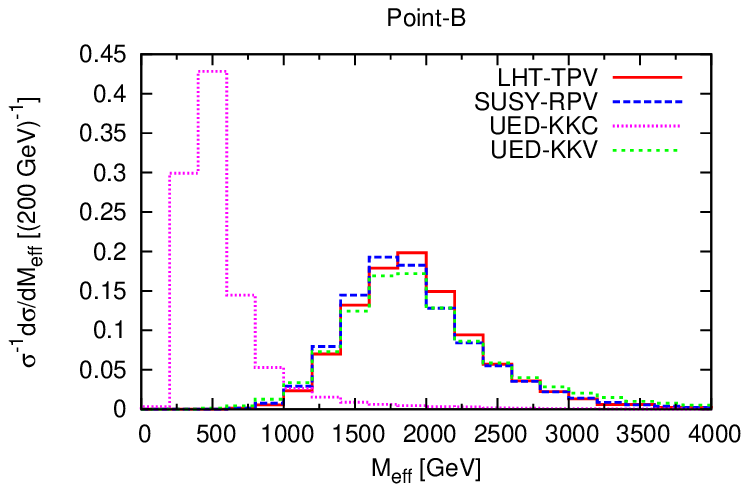,width=8.5cm,height=7.0cm,angle=-0}} 

\caption{Normalized effective mass distribution in various models for point A
  (left figure) and point B (right figure), at $\sqrt{s}=14 \tev$
  after Cut-2.}
\label{fig:Meff}
\end{center}
\end{figure}


Note that although usually $M_{eff}$ carries the information about the
mass scale of the particles produced in the initial parton level hard
scattering, this is not true in all cases. Specifically, it is
well-known that for a mass spectrum which is very closely spaced,
$M_{eff}$ largely underestimates the relevant mass-scale, the reason
being very similar to that for the MET distribution being low in
UED-KKC.

If an excess is seen in the $4l+nj+\met$ ($n\geq2$) channel, fingers
can be pointed at the several models mentioned before. Figure~\ref{fig:MET} is an example where the
MET distributions show similar behaviour when the strongly interacting
heavy particles in all of the four aforementioned models under
consideration have similar masses. We thus conclude that all the four
models clearly qualify as missing energy look-alikes. The minor
differences that exist are difficult to use as discriminators, as
these can be masked by features of the detector as well as systematic
errors. In Figure~\ref{fig:Meff}, we also present the $M_{eff}$
distributions for all these models, for the same `mass scale' of $\sim
600 \gev~(1000 \gev)$ for point A (point B). One finds that all the
models except UED-KKC have a peak in the $M_{eff}$ distribution at
around twice the mass-scale (i.e., $\sim 1200 \gev$ for point A). For
point A in UED-KKC, the distribution peaks at around 300 GeV. Similar
features are also observed for point B.

This fact, if correlated with the total cross-section, can perhaps be
used to single out the UED-KKC model from the remaining
three. However, one can still vary the excited quark and gluon masses
in UED-KKC to match both the MET distributions for other models with a
600 GeV mass-scale. In order for this to happen, however, $R^{-1}$ has
to be as large as about 3 TeV. This would not only lead to very low
total cross-section but also imply a scenario that is ruled out due to
overclosure of the universe through $\gamma_1$.

However, while similar MET distribution but much softer $M_{eff}$
distribution can single out UED-KKC, similar distributions in both
variables but very low cross-sections can be noticed in special
situations in the other models as well. For example, one can have RPV
SUSY with both the $\lambda$-and $\lambda^{\prime}$-type interactions,
with the later being of larger value, thus suppressing the decay of
the lightest neutralino into two leptons. In such a case, with
$m_{\tilde q} \simeq m_{\tilde g} \simeq$ 600 GeV, the MET as well as
$M_{eff}$-distributions will be very similar as earlier, but the
cross-section may be down by a large factor.

The above example shows that there is a pitfall in using total rate as
a discriminant.  Keeping this in view, we choose our benchmark points
with similar masses for all the models but look for other kinematical
features where the differences show up.

\section{Analysis of $4l$ events}
\label{4l-detail}
\subsection{Four-lepton invariant mass distribution: two classes}
In the table containing rates of four-lepton final states, we have
allowed all the leptons to be of either charge. While we keep open the
option of having all charge combinations, in this and the next two
subsections we concentrate on those events where one has two positive
and two negatively charged leptons. The issue of `total charge of the
set of leptons' and its usefulness in model discrimination is taken up
section 5.4.

\begin{figure}[h!]
\begin{center}
\centerline{\epsfig{file=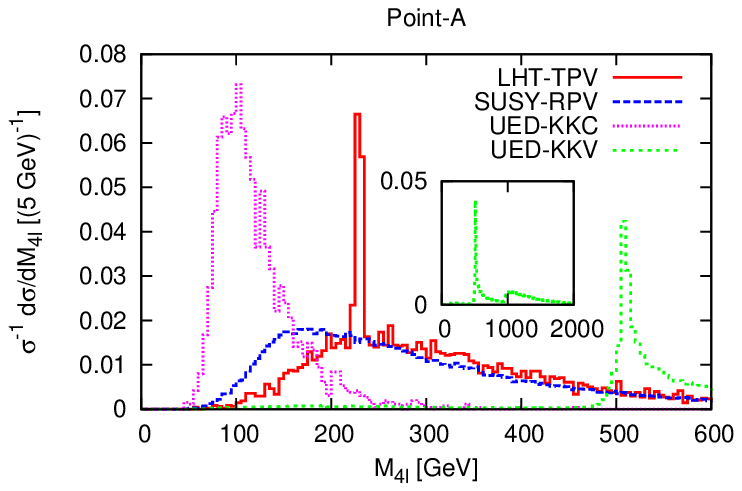,width=8.5cm,height=7.0cm,angle=-0}
\epsfig{file=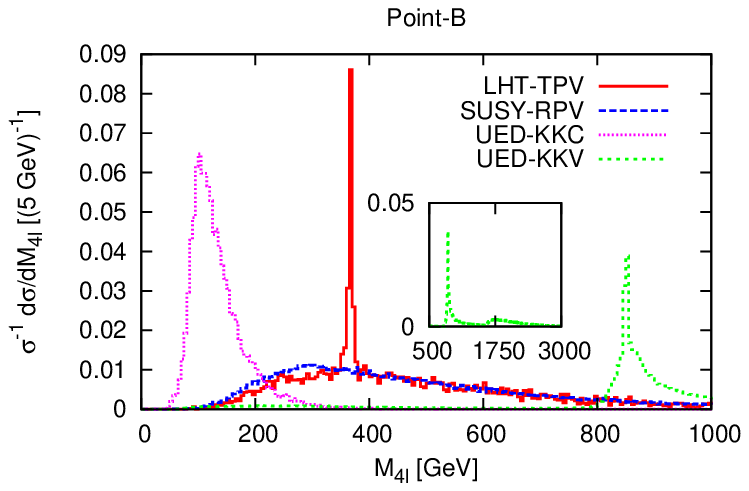,width=8.5cm,height=7.0cm,angle=-0}} 

\caption{Normalized invariant mass distribution of four leptons
  ($M_{4l}$) in the different models for point A (left figure) and
  point B (right figure), at $\sqrt{s}=14 \tev$. The insets show the
  full range of $M_{4l}$ for UED-KKV. The distributions have been
  plotted after Cut-2.}
\label{fig:IM4l}
\end{center}
\end{figure}

For the $4l+nj+\met$ channel (with $n\geq2$), the first
distribution that we look into is that of the four-lepton invariant
mass ($M_{4l}$). From Figure~\ref{fig:IM4l} we can see that, based on
the $M_{4l}$ distribution the models can be classified into two
categories, 1) those with a peak in the $M_{4l}$ distribution and 2)
those without any such peak. 

In SUSY-RPV, there is no single bosonic particle decaying, via cascade,
to four leptons. Hence, no invariant mass peak is expected in this
model and the $M_{4l}$ distribution has a very broad shape, as we see
in Figure~\ref{fig:IM4l}. As discussed earlier, in the example taken,
the four leptons here come from the RPV decays of the neutralino and
the RPC decays of the chargino.

Similarly, in UED-KKC the four leptons come from the two cascade decay
chains and not from a single particle. In this case, the decay of a
$Z_1$ gives two leptons of opposite charge and same flavour in each
chain. Note that the range in which $M_{4l}$ takes values for UED-KKC
is rather restricted, as can be seen in Figure~\ref{fig:IM4l}. We can
understand this in the following manner. Let us denote the
four-momenta of the two leptons coming from one chain by $p_1$ and
$p_2$, and those of the other two from the second chain by $p_3$ and
$p_4$. Then, an approximate upper bound on $M_{4l}$ can be obtained as
follows: \bea M_{4l}^2&=&(p_1+p_2+p_3+p_4)^2
\nonumber\\ &=&(p_1+p_2)^2+(p_3+p_4)^2+2(p_1+p_2).(p_3+p_4) \eea Now,
\bea (p_1+p_2)^2\leq
\frac{{(M_{Z_1}^2-M_{L_1}^2)}{(M_{L_1}^2-M_{\gamma_1}^2)}}{M_{L_1}^2}
\eea A similar bound is also applicable to $(p_3+p_4)^2$. Finally, we
can approximate \bea 2p_1.p_3=2(E_1E_3-\vec{p_1}.\vec{p_3}) \simeq
2E_1E_3(1-cos\theta_{13}) \leq 4E_1E_3 \eea Here $\theta_{13}$ is the
angle between $\vec{p_1}$ and $\vec{p_3}$. Hence, we can finally
approximate the upper bound on $M_{4l}$ for UED-KKC as \be M_{4l}\leq
\sqrt{2\frac{{(M_{Z_1}^2-M_{L_1}^2)}{(M_{L_1}^2-M_{\gamma_1}^2)}}{M_{L_1}^2}+
  4[E_1E_3+E_1E_4+E_2E_3+E_2E_4]} \ee
 
The first term within the square-root has the numerical value of about
$2400 {\gev}^2$ for point A in UED-KKC. Because of the unknown boost
of the partonic centre of mass frame, we cannot put any definite upper
bound on the $p_T$'s of $Z_1$ and $L_1$, and therefore the maximum
possible values of the lepton energies $E_i$ are also undetermined.
Still, as very little energy is available at the rest frame of the
decaying KK-excitations (once again because of low mass-splittings),
$E_i$ cannot be very large. This is the reason we expect $M_{4l}$ for
UED-KKC to take values in a somewhat restricted range.

For UED-KKV, we observe a very sharp peak in the $M_{4l}$ distribution
around $\sim 510 \gev~(850 \gev)$ which is the mass of $Z_1$ for point
A (point B). The four leptons in this case come from the cascade decay
of a $Z_1$ to a lepton pair and a $\gamma_1$ followed by the KKV decay
of the $\gamma_1$ to another pair of leptons. There is of course a
long tail in the $M_{4l}$ distribution here coming from the cases
where all four of the leptons do not come from the decay of a single
$Z_1$. For example, they can come from the leptonic decay of the two
$\gamma_1$'s via KK-parity violating interactions. We demonstrate this
full range in the insets of Figure~\ref{fig:IM4l}. It is, however,
important to note that, in these cases, $M_{4l}$ mostly exceeds
$M_{\gamma_1}$ when two of the leptons come from $\gamma_1$ decay and
the two others come from the cascade. On the other hand, the invariant
mass is greater than $2M_{\gamma_1}$ when the four leptons come from
the decay of two $\gamma_1$'s. This is why we observe an excess after
around $1 \tev ~ (1.6 \tev)$ for point A (point B) in the insets of
Figure~\ref{fig:IM4l}.  As long as we have a $\gamma_1$ LKP with
KK-parity violating interactions, this very clear peak, followed by a
hump, observable in the $M_{4l}$ distribution is a very important
feature of UED-KKV.

Finally in LHT-TPV, we see in Figure~\ref{fig:IM4l} a very sharp peak
superposed in a broad overall distribution. The reason for this is
that in the example taken, most of the four-lepton events are due to
the decay of an $A_H$ to $W^+W^-$ followed by the leptonic decays of
the $W$'s. These events will not give rise to any invariant mass
peak. But, there is a fraction of events where one $A_H$ decays to a
$ZZ$ pair which then in turn can decay to four-leptons. This will give
rise to a peak in $M_{4l}$ as we see in Figure~\ref{fig:IM4l}. Note
that, the branching fraction of an $A_H$ of mass 230 GeV to a pair of
Z-bosons is around 22\% while it is 77\% to a pair of W's. Over and
above that, the branching fraction of $W^\pm$ to charged leptons is
greater than that of Z. This accounts for the spread of events away
from the peak, as seen in Figure~\ref{fig:IM4l}. One may also note
that, in general, fewer 4-lepton events are expected in this model
from leptonic cascades of the $Z_H$. This is because, for our choice
of parameters, $Z_H$ has the largest branching ratio in the $h A_H$
channel.

Based on the above considerations, LHT-TPV and UED-KKV fall into
category-1 while SUSY-RPV and UED-KKC belong to category-2. Note that,
this classification for LHT-TPV is valid only for $A_H \ge 120$ GeV,
so that it has sufficient branching fraction for $ZZ^{(*)}$. We shall
discuss the other cases later in subsection~\ref{subsec:invmass-2l}.

Our task now is to distinguish between the look-alike models within
each category. We do so in the following subsections.

\subsection{Pairwise dilepton invariant mass distribution of the four leptons}

One can go a little further in the task of model discrimination by
combining the information from $M_{4l}$ plots with distributions in
the invariant mass distribution
$M_{ij}$ ($i,j=1,2 $) of the OS lepton pairs. Taking events with two pairs
of opposite-sign leptons, we first order the leptons of each sign in
the descending order of transverse momentum.  In order to be
sufficiently general in our analysis, no constraint on the flavour of
the leptons is imposed while pairing them up. This is because, in 
SUSY-RPV, flavours of the leptons coming from neutralino decays are not
correlated in general, apart from constraints coming from the
antisymmetry in the first two indices of the $\lambda_{ijk}$-type
terms in the superpotential. Thus in general we can form four possible
pairs of opposite sign (OS) leptons.

\begin{figure}[h!]
\begin{center}

\centerline{\epsfig{file=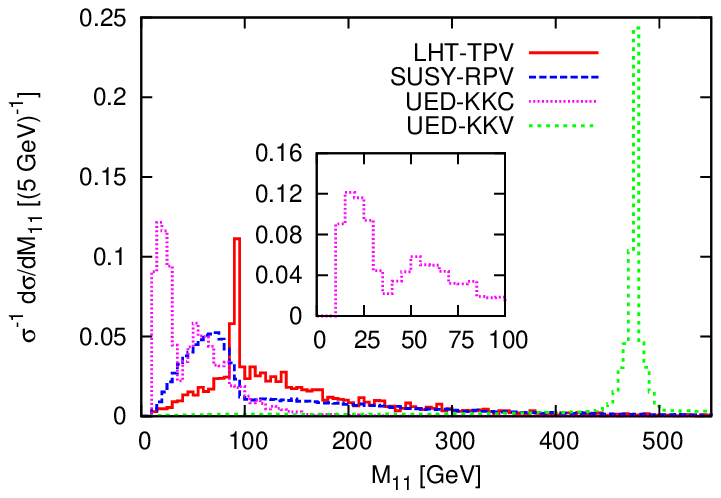,width=8.5cm,height=6.85cm,angle=-0}
\hskip 20pt \epsfig{file=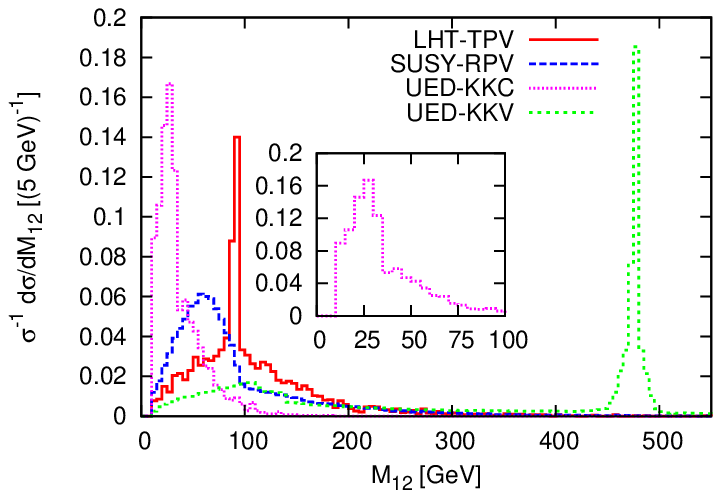,width=8.5cm,height=6.85cm,angle=-0}}

\centerline{\epsfig{file=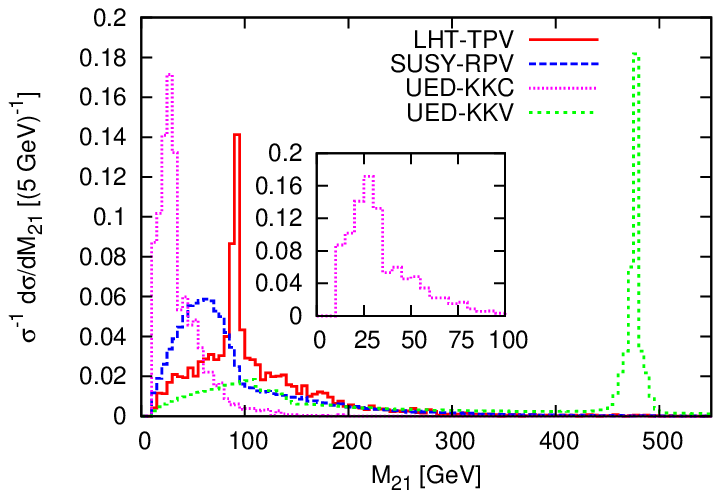,width=8.5cm,height=6.85cm,angle=-0}
\hskip 20pt \epsfig{file=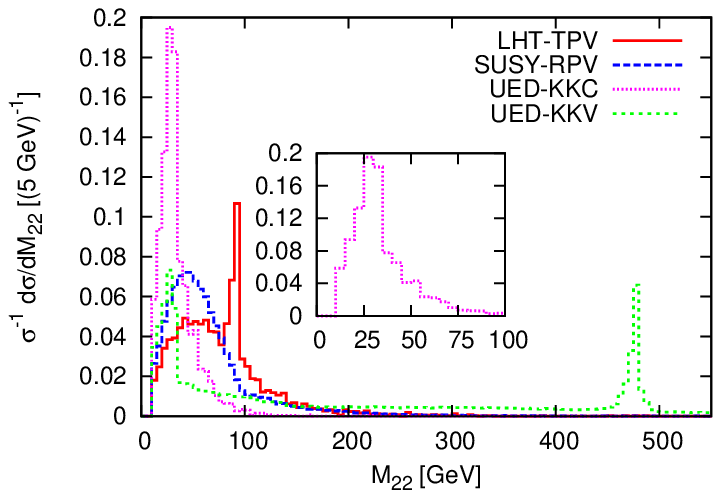,width=8.5cm,height=6.85cm,angle=-0}}

\caption{Normalized invariant mass distribution of OS lepton pairs
  formed out of the four leptons with total charge zero. The above
  distributions are for point A, at $\sqrt{s}=14 \tev$, after
  Cut-2. The insets are shown in order to clearly identify the
  position of the invariant mass edge for UED-KKC.}
\label{fig:2l-14TeV-600}
\end{center}
\end{figure}

For the models belonging to {\it category-2}, i.e., SUSY-RPV and
UED-KKC, the $M_{ij}$ distributions show no peaking behaviour. But, we
see in the insets of Figure~\ref{fig:2l-14TeV-600} that the $M_{ij}$
distributions for UED-KKC show a prominent edge at $\sim 35~(50) \gev$
for point A (point B). We note that in 50\% of the cases,
the lepton pairs come from the same decay chain starting with a
$Z_1$ (as described in detail in the context of angular correlation between the lepton-pairs in
subsection~\ref{subsec:angular}). Therefore, an invariant mass edge is expected at the value
given by
$\sqrt{{{(M_{Z_1}^2-M_{L_1}^2)}{(M_{L_1}^2-M_{\gamma_1}^2)}}/{M_{L_1}^2}}$,
which comes out to be $32.36 ~(50.95) \gev$ for point A (point B).

\begin{figure}[h!]
\begin{center}

\centerline{\epsfig{file=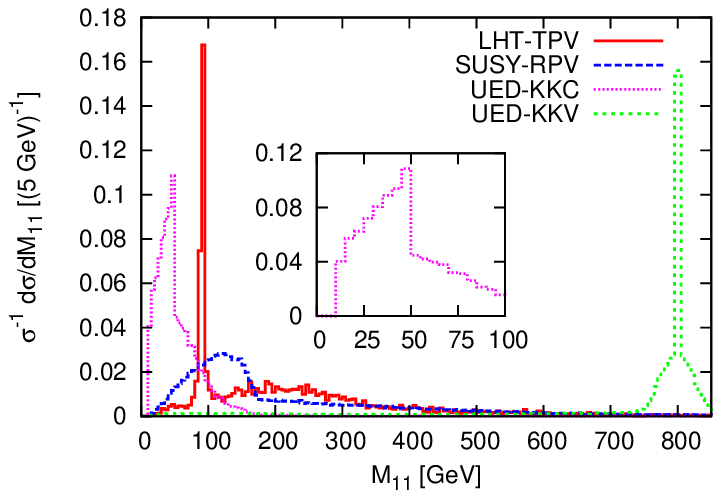,width=8.5cm,height=6.85cm,angle=-0}
\hskip 20pt \epsfig{file=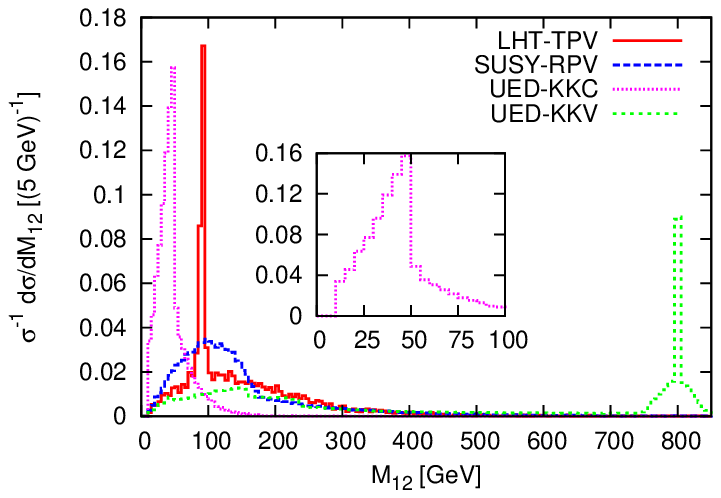,width=8.5cm,height=6.85cm,angle=-0}}

\centerline{\epsfig{file=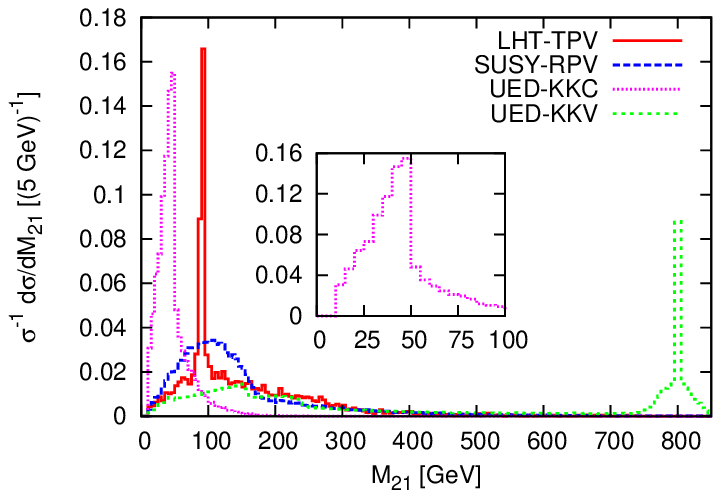,width=8.5cm,height=6.85cm,angle=-0}
\hskip 20pt \epsfig{file=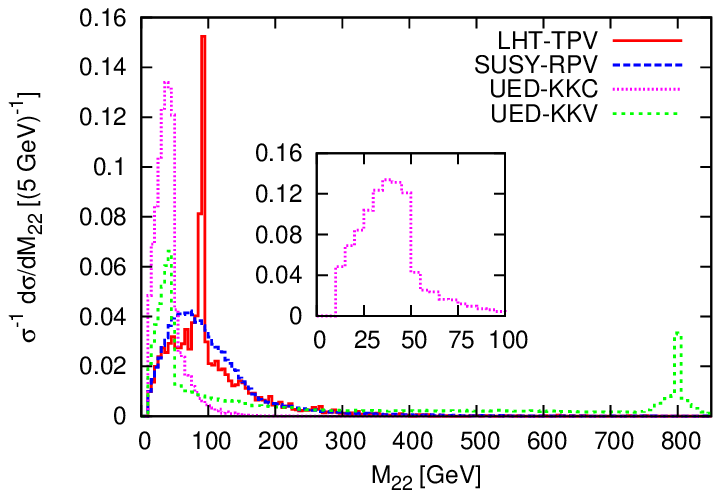,width=8.5cm,height=6.85cm,angle=-0}}
\caption{Same as Figure~\ref{fig:2l-14TeV-600}, for point B.} 
\label{fig:2l-14TeV-1000}
\end{center}
\end{figure}

For the SUSY-RPV point under consideration, we also see in
Figures~\ref{fig:2l-14TeV-600} and~\ref{fig:2l-14TeV-1000} an edge in
the invariant mass distributions of two OS leptons near the
corresponding neutralino mass. But as the two leptons in the pairs we
consider have equal probability of coming from either the same chain
or two different chains, the mass-edge in the distributions is smeared
by the ``wrong combinations''. If one forms OS leptons pairs with a
very small opening angle between them, the neutralino mass-edge will
become very sharp. As the lightest neutralino mass can be as low as
$46 \gev$~\cite{PDG}, the position of the invariant mass edge for
UED-KKC and SUSY-RPV will not be very different for neutralino masses
close to this bound. However, for neutralino masses higher than $\sim
100 \gev$, the edge positions will be quite different, as in UED-KKC
the value obtained cannot be that large in any allowed region of the
parameter space.

In case of models belonging to {\it category-1}, we can see from
Figures~\ref{fig:2l-14TeV-600} and~\ref{fig:2l-14TeV-1000} that the
distributions for UED-KKV show a clear peak at the same value of
$M_{ij}$ for all the four possible cases. The peak is at the mass of
$\gamma_1$ which is $475 ~(800) \gev$ for point A (point B). The two
OS leptons coming from the decay of a $\gamma_1$ are expected to be of
high transverse momentum, thereby constituting the majority of (11)
pairs. Therefore, the peak height is very large for the $M_{11}$
distribution. As we gradually consider OS combinations of softer
leptons, the peak height reduces . This is because, most of the softer
leptons come from intermediate stages of the cascade, and invariant
mass distributions involving them will just give rise to a broad
overall distribution. Therefore, in UED-KKV, as long as the LKP is the
heavy photon, such an invariant mass peak is an unmistakable feature
of the model. Here, one should also note that the position of the OS
dilepton peak is always different from the position of the four-lepton
peak as we obtain in UED-KKV. The difference, although not very large,
can be observed at the LHC given the fact that the four-lepton channel
is a rather clean one. Looking at Figures~\ref{fig:IM4l} and
\ref{fig:2l-14TeV-600} we find that for point A, the $M_{4l}$ peak is
roughly at 509 GeV ($M_{Z_1}$) and the $M_{ij}$ peaks are at 475 GeV
($M_{\gamma_1}$).  Hence the difference between the values of these
two peak positions, which in this case is around 34 GeV, is large
enough to be measurable within the experimental resolutions at the
LHC. Moreover, note that, these two peaks appear in two different
distributions making the resolution even easier. Similar
considerations apply for point B, too.

In the example considered here, the corresponding distributions for
LHT-TPV also show a clear peak at the mass of the Z-boson ($\sim 90
\gev$), as we can see from Figures~\ref{fig:2l-14TeV-600}
and~\ref{fig:2l-14TeV-1000}. The Z-peak is superposed over a broad
$M_{ij}$ distribution coming from leptons from W's. Note that for
point A, the branching fraction of $A_H\rightarrow ZZ$ is 22\% and the
corresponding branching fraction $A_H\rightarrow W^+W^-$ is
77\%. Therefore, relatively fewer dilepton pairs come from the decay
of a Z-boson and hence the fraction of events under the Z-peak is
smaller compared to the total distribution. For point B, as the
$A_H\rightarrow ZZ$ branching fraction increases to 35\%, the peak at
$M_Z$ is even more prominent over the $l^+l^-$ continuum coming from
$W^+W^-$. This, however, is not generic to LHT-TPV. As discussed in
detail in Ref.~\cite{Freitas_Schwaller_Wyler,Mukhopadhyay}, as far as
the decays of $A_H$ is concerned, there are essentially three possible
cases for this model.  For lower masses ($M_{A_H} \lapprox 120 \gev, f
\lapprox 800 \gev$) the decay of $A_{H}$ is dominated by the
loop-induced two-body modes into fermions (case-1), whereas for higher
masses ($M_{A_H} > 2M_{W} , f > 1070 \gev$) the two-body modes to
gauge boson pairs dominate (case-3).  For intermediate masses, both
the two-body and three-body modes compete (in the three-body mode we
have one off-shell $W^{\pm}$ or $Z$) (case-2). The points we have
analysed here (point A and B), belong to case-3. For case-1 one would
obtain a very clear peak in the OS dilepton invariant mass
distribution but no peak in $M_{4l}$, while in case-2 a peak is
expected in both OS dilepton and four-lepton distributions at the same
mass-value. Therefore, case-1 will give rise to a situation where
LHT-TPV belongs to category-2, but it will be distinguishable from
SUSY-RPV and UED-KKC by its dilepton peak. And in case-2, LHT-TPV
would belong to category-1, but it can still be distinguishable from
UED-KKV by the fact that the OS dilepton and four-lepton peaks will be
at the same mass-value for LHT-TPV, but at different values for
UED-KKV. Moreover, in case -2, $120 \gev \lapprox M_{A_H} \lapprox 160
\gev$, and no peak is expected in UED-KKV at such a low value of the
dilepton invariant mass. For details on the reconstruction of such
peaks above SM backgrounds in LHT-TPV, we refer the reader to
Ref.~\cite{Mukhopadhyay}. Thus, we note that the
features of OS dilepton invariant mass distributions, if used in
conjunction with other important kinematic characteristics, can help
in discriminating models belonging to different categories.

\begin{figure}[h!]
\begin{center}
\epsfig{file=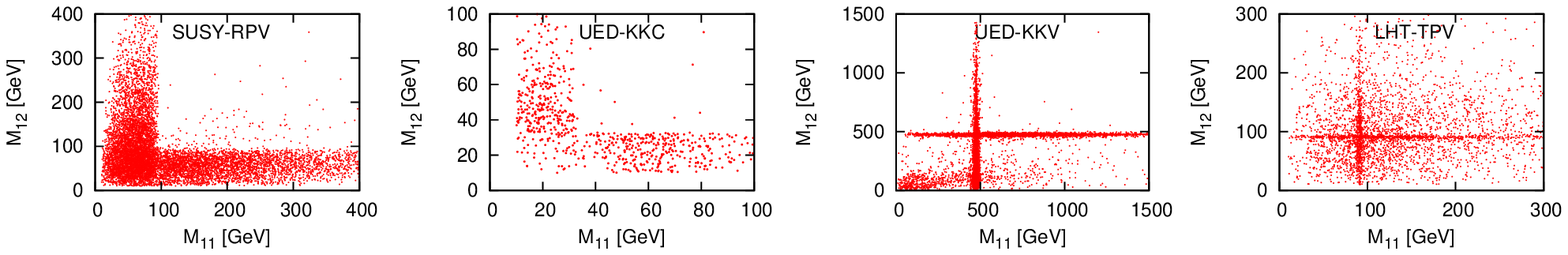,width=17cm,height=4cm,angle=-0}\\
\epsfig{file=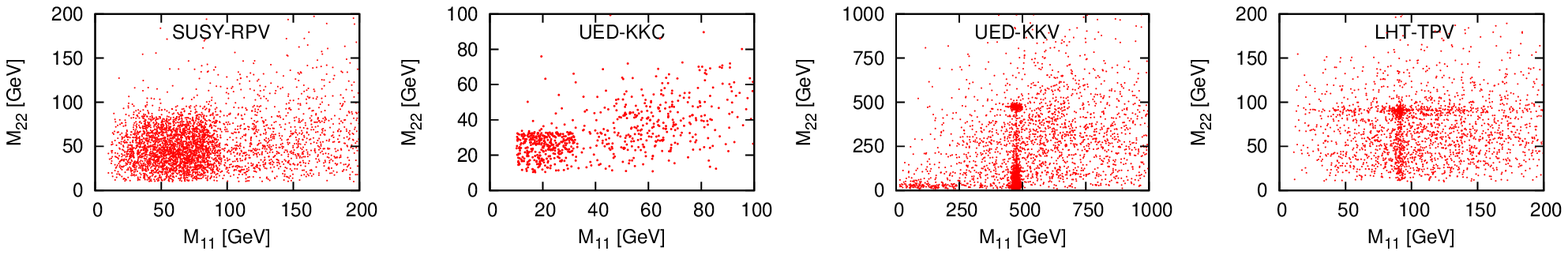,width=17cm,height=4cm,angle=-0}

\caption{Pairwise correlation of opposite-sign dilepton invariant mass on an event-by-event basis for point A in various models at $\sqrt{s}=14 \tev$, after Cut-2. We have shown two of the six possible correlations, the essential features in the rest being the same.}
\label{fig:correlation}
\end{center}
\end{figure}

In addition to the invariant mass distributions of the different possible OS dilepton pairs formed out of the four leptons, it might also be useful to look into pairwise correlation of the different $M_{ij}$'s as shown in Figure~\ref{fig:correlation}. In this figure, for the purpose of illustration, we show the correlation of ($M_{11},~M_{12}$) and ($M_{11},~M_{22}$) for the various models at point A on an event-by-event basis (at $\sqrt{s}=14 \tev$ after Cut-2). We can see from Figure~\ref{fig:correlation} that for the models where we observe peaks in the $M_{ij}$ distributions, the ($M_{11},~M_{12}$) correlation plot shows two typical lines parallel to the $M_{11}$ and $M_{12}$ axes at the values of the peak-position (around $475 \gev$ for UED-KKV and $90 \gev$ for LHT-TPV). On the otherhand, for the models showing an edge in the $M_{ij}$ distributions we see from the scatter plots that the values of $M_{ij}$ are mostly concentrated within the upper-bound given by the position of the edge (around $100 \gev$ for SUSY-RPV and $35 \gev$ for UED-KKC). 

If both the OS dilepton pairs $11$ and $22$ come from a $\gamma_1$ in UED-KKV or a $Z$-boson in case of LHT-TPV, then $M_{11}$ and $M_{22}$ will have the same value (with an error-bar, due to detector effects etc.) in the corresponding event. That's why we expect a ``blob'' in the ($M_{11},~M_{22}$) plane as can be seen from Figure~\ref{fig:correlation} for UED-KKV and LHT-TPV (centered around $M_{\gamma_1}$ and $M_Z$ respectively). For UED-KKC or SUSY-RPV, the boundedness of $M_{ij}$'s is once again reflected by the ``box-like'' distribution, the spread away from the boxes being significantly lesser than for the ($M_{11},~M_{12}$) case, as the leptons in the pair $22$ are much softer. As we have four possible OS lepton pairings as explained before, one can have six different correlation plots between them. But all the essential features remain the same as in the two correlation plots we show for the various models.

\label{subsec:invmass-2l}

\subsection{Pairwise angular correlation of the four leptons}
In this subsection, we consider the correlation in the opening angle between the lepton pairs. The pairs have been formed out of the
four leptons following the same prescription as used in the previous
subsection.

\begin{figure}[h!]
\begin{center}
\centerline{\epsfig{file=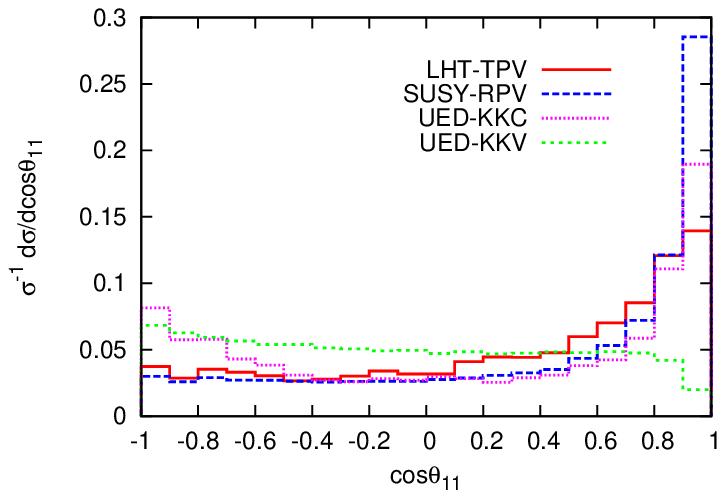,width=8.5cm,height=6.85cm,angle=-0}
\hskip 20pt \epsfig{file=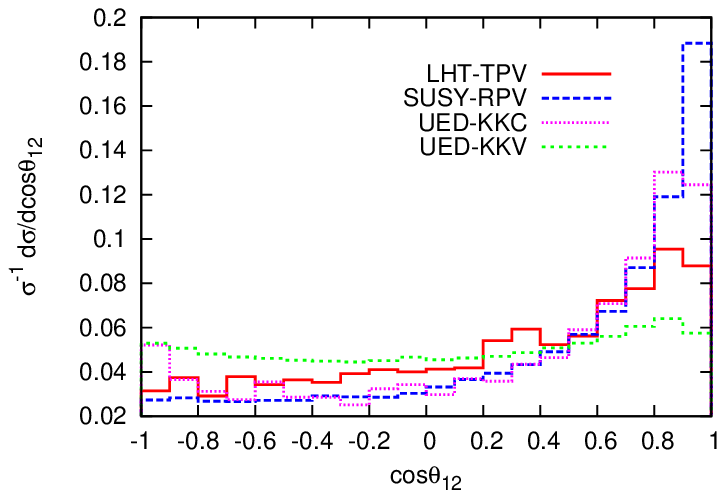,width=8.5cm,height=6.85cm,angle=-0}}
\centerline{\epsfig{file=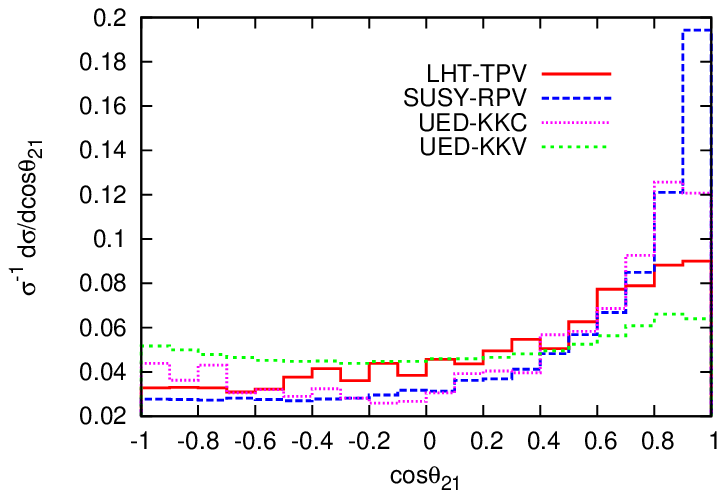,width=8.5cm,height=6.85cm,angle=-0}
\hskip 20pt \epsfig{file=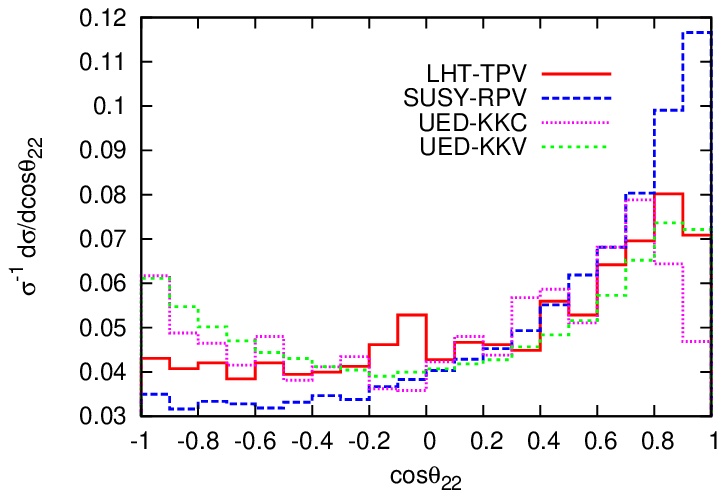,width=8.5cm,height=6.85cm,angle=-0}}
\caption{Normalized distribution of cos$\theta_{ij}$ in the different
  models for point A at $\sqrt{s}=14 \tev$ after Cut-2.}
\label{fig:cos-14TeV-600}
\end{center}
\end{figure}

\begin{figure}[h!]
\begin{center}
\centerline{\epsfig{file=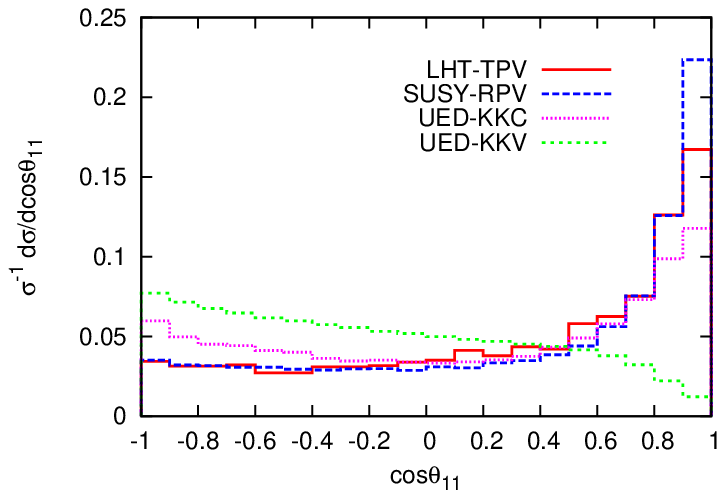,width=8.5cm,height=6.85cm,angle=-0}
\hskip 20pt \epsfig{file=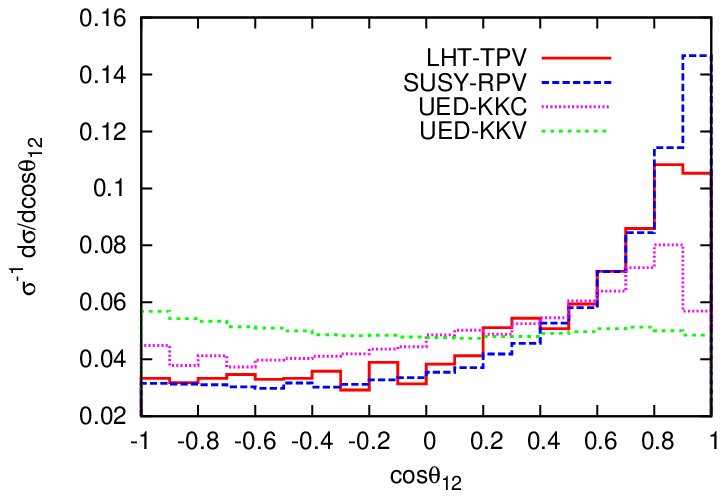,width=8.5cm,height=6.85cm,angle=-0}}
\centerline{\epsfig{file=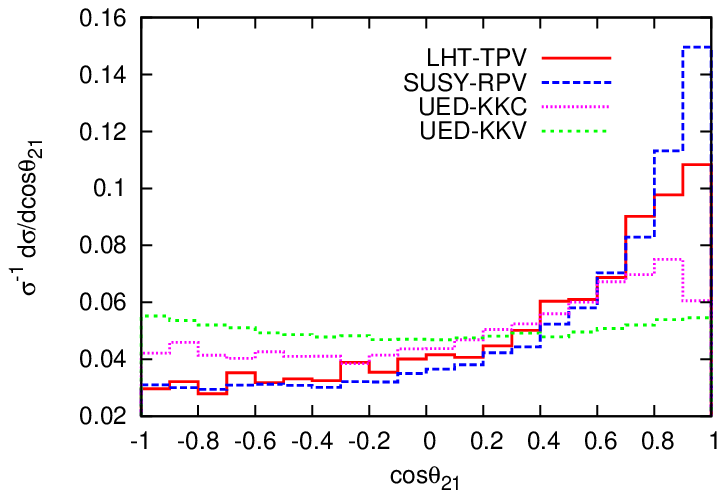,width=8.5cm,height=6.85cm,angle=-0}
\hskip 20pt \epsfig{file=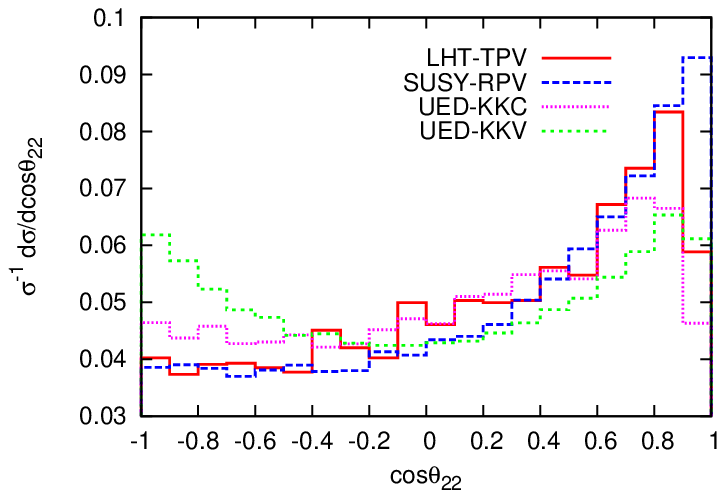,width=8.5cm,height=6.85cm,angle=-0}}
\caption{Normalized distribution of cos$\theta_{ij}$ in the different
  models for point B at $\sqrt{s}=14 \tev$ after Cut-2.}
\label{fig:cos-14TeV-1000}
\end{center}
\end{figure}

In Figures~\ref{fig:cos-14TeV-600} and~\ref{fig:cos-14TeV-1000} we
have presented distributions of the cosine of the opening angle
between the OS lepton pairs for point A and point B respectively. The
opening angle is calculated in the lab frame and is defined by the
relation \be
cos\theta_{ij}=\frac{\vec{p_i}.\vec{p_j}}{|\vec{p_i}||\vec{p_j}|}~;
~~~i,j=1,2 \label{cos} \ee where $\vec{p_i}$ and $\vec{p_j}$ stand for
the momenta of positive and negatively charged leptons respectively.

In order to quantify the difference between the various models in the
distribution of $cos\theta_{ij}$, we define the following asymmetry
variable: \be
A_{ij}=\frac{{\int_0}^{+1}\frac{d\sigma}{dcos\theta_{ij}}
  d(cos\theta_{ij})-{\int_{-1}}^{0}\frac{d\sigma}{dcos \theta_{ij}}
  d(cos\theta_{ij})}{{{\int_0}^{+1}\frac{d\sigma}{dcos\theta_{ij}}d(cos\theta_{ij})+
    {\int_{-1}}^{0}\frac{d\sigma}{dcos\theta_{ij}}d(cos\theta_{ij})}}. \label{asy}
\ee As we are considering the normalized distribution of
$cos\theta_{ij}$ here, the denominator of $A_{ij}$ as defined in
Equation~\ref{asy} will be $1$. In Table~\ref{tab:asymmetry} we show
the values of $A_{ij}$ in the different models under consideration.

Let us first consider the {\it models belonging to category-2}. We can
see from Figures~\ref{fig:cos-14TeV-600} and~\ref{fig:cos-14TeV-1000}
that for SUSY-RPV, the distribution of $cos\theta_{ij}$ tends to peak
towards $+1.0$. This is because the fraction of OS lepton pairs which
are coming from the decay of a sufficiently boosted single neutralino
tend to be highly collimated. The rest of the combinations (which
include two leptons coming from two separate decay chains, or one
coming from a chargino and another from a neutralino decay) will have
OS leptons which are largely un-correlated. Thus we can see a overall
flat distribution of $cos\theta_{ij}$ with a very significant peaking
towards $+1.0$. For the same reason, the values of $A_{ij}$ in
SUSY-RPV in Table~\ref{tab:asymmetry} are larger than in the other
models for all $i,j$. We should note here that the boost of the
neutralino is large, thanks to the substantial mass splitting among
the gauginos usual in mSUGRA. In a generic MSSM model, the splittings
might become smaller, thereby rendering the neutralinos less boosted
and the distributions of $cos\theta_{ij}$ less sharply peaked towards
$+1.0$.

For UED-KKC, we have two possibilities while forming the OS lepton
pairs - the two leptons can either come from the same decay chain or
they can come from two separate decay chains. In the former case, we
expect the two OS leptons to have a smaller opening angle in
general. This is because, they come from the decay chain
$Z_1\rightarrow l^+ l^- \gamma_1$, where the parent $Z_1$ will be
carrying the boost of the initially produced $Q_1$ . As the leptons
themselves have very low $p_T$'s in the rest frame of the $Z_1$, this
boost of the $Z_1$ will make them collimated to an extent. For the
second possibility, the leptons are, in many cases, nearly
back-to-back as they come from the cascade decay of two $Z_1$'s which
for a significant fraction of events lie in two different hemispheres
of the detector. For point A, the mass splittings are quite small
between the $n=1$ KK-excitations, whereas for point B, they are
relatively larger. Since we have demanded two of the leptons to have
$p_T$ greater than $20 \gev$, it is very likely from the point of view
of UED mass splittings that for point A, in a larger fraction of
events, both of the hard leptons will not come out from the same decay
chain . Therefore, in a significant number of events, the combinations
(12) and (21) are expected to be from the same chain, whereas the (11)
and (22) pairs will involve both the decay chains. This leads us to
expect the cos$\theta_{12}$ and cos$\theta_{21}$ distributions to be
more peaked towards $+1.0$ than towards $-1.0$, while the
cos$\theta_{11}$ and cos$\theta_{22}$ distributions have slight
peaking near both $+1.0$ and $-1.0$. It can be readily verified that
Figure~\ref{fig:cos-14TeV-600} is in conformity with these
observations.  Similarly, we see from Table~\ref{tab:asymmetry} that
(for point A) the values of $A_{12}(=0.34)$ and $A_{21}(=0.34)$ are
relatively higher and positive, implying significant asymmetry with
more events having cos$\theta>0$. We also note that $A_{11} (=0.17)$
and $A_{22}=(0.12)$ indicate more symmetric distributions for
cos$\theta_{11}$ and cos$\theta_{22}$ as discussed above. This
asymmetry, however, is no longer expected if the mass-splittings
become larger, as then the OS lepton pairs can come with equal
probability from both the chains. We observe this feature in case of
point B in Figure~\ref{fig:cos-14TeV-1000} and also in
Table~\ref{tab:asymmetry}. Therefore, in UED-KKC, the distribution of
cos$\theta_{ij}$ is in general {\em much less asymmetric towards
  $+1.0$} as compared to SUSY-RPV.

Thus, based upon the above features, one should be able to distinguish
between SUSY-RPV and UED-KKC. We shall see in what follows that there
are also other features like the total charge of the four leptons, or
the ratio of five or higher lepton to four-lepton cross-sections which
can act as useful discriminators between the models belonging to
category-2.

Next we consider the {\it models belonging to category-1}.  In case of
UED-KKV, the two OS leptons of highest $p_T$ come in most cases from
the decay of a $\gamma_1$ which itself has a very low $p_T$ to start
with. So, in the rest frame of $\gamma_1$, the two leptons will almost
be back-to-back, leading to the slight peaking of $cos\theta_{11}$
towards $-1.0$ as seen from Figures~\ref{fig:cos-14TeV-600}
and~\ref{fig:cos-14TeV-1000}. This interesting feature of UED-KKV is
also reflected in the value of $A_{11}=-0.11 ~(-0.24)$ for point A
(point B). The distributions for cos$\theta_{12}$ and cos$\theta_{21}$
on the other hand are very flat indicating that these leptons have no
angular correlation between them. This is not unexpected, since the
combinations (12) and (21) will generally be formed when one of the
leptons (the harder one with index 1) comes from the decay of
$\gamma_1$, and the other, softer one, from the intermediate stages of
the cascade (mostly from the decay of $Z_1$, $W_1$ or
$L_1$). Therefore, $A_{12}$ and $A_{21}$ are close to zero as can be
seen from Table~\ref{tab:asymmetry}. Finally, we see that the
distribution for cos$\theta_{22}$ shows a peaking behaviour towards
both $+1.0$ and $-1.0$. This can be understood from the fact that the
two softest leptons almost always come from the intermediate stages of
the cascade where they emerge from either the decays of $Z_1$ and
$W_1$. A large fraction of $Z_1$-initiated events gives rise to
collimated leptons (giving rise to the peaking towards $+1.0$, while a
significant fraction of events where the leptons come from $W_1$ will
have them coming out in opposite directions (responsible for the
peaking towards $-1.0$.

Finally, from all the distributions of cos$\theta_{ij}$ and the values
of $A_{ij}$ it is clear that the LHT-TPV model has features similar to
SUSY-RPV. In particular, as the leptons come from the decays of a
boosted $A_H$ (via intermediate $W^+W^-$ or $ZZ$ states), they tend to
be collimated. The collimation here is somewhat less than as in
SUSY-RPV because of the fact that for the chosen parameters (point A),
the squark (gluino)-neutralino mass difference ($\sim 471(509) \gev$)
is much larger than the $q_H$-$A_H$ mass difference ($\sim 375 \gev$).

Thus while both UED-KKV and LHT-TPV might give rise to clear peaks in
the $M_{4l}$ distribution, they show {\em entirely different
  behaviour} as far as the angular correlation between the OS lepton
pairs is concerned. This, therefore, can act as a very good
discriminator between these models belonging to category-1. In the
following subsections, we shall look into some more variables that can
be used to further clarify the distinction between these two models,
especially when different types of mass-spectra in LHT-TPV tend to
obliterate the $4l$ mass peaks so spectacular for our chosen benchmark
points.

We should note here that there is a dip observable in the
cos$\theta_{ij}$ distributions towards the last bin near $+1.0$ for
UED-KKC and LHT-TPV. This dip, however, is not seen in the
cos$\theta_{11}$ distribution. This is stemming from the fact that we
have demanded all OS lepton pairs to have an invariant mass greater
than 10 GeV. Now, when the lepton $p_T$'s are not very high
themselves, the invariant mass becomes very small when the angle
between the leptons tends to zero. These events, therefore, have been
removed by the above cut, giving rise to the observed dip. As for the
(11) combination the leptons are much harder, the lepton pairs always
pass the invariant mass cut, and this feature does not appear for this
case.

\label{subsec:angular}
\begin{table}[htb]
\begin{center}
\begin{tabular}{||c|c|c|c|c||}
\hline \hline
\multicolumn{5}{||c||}{Point A}\\\hline\hline
Variable & SUSY-RPV & LHT-TPV & UED-KKC & UED-KKV \\ 
\hline
$A_{11}$    &  0.46    &  0.37  &  0.17   & -0.11    \\
\hline 
$A_{12}$    & 0.44     &   0.28 &  0.34   & 0.06        \\
\hline
$A_{21}$    & 0.44     &   0.28 &  0.34   & 0.07        \\
\hline
$A_{22}$    & 0.32     &   0.15 &  0.12   & 0.08     \\

\hline \hline 
\multicolumn{5}{||c||}{Point B} \\\hline\hline
$A_{11}$    &  0.39    &  0.38  &  0.16   & -0.24    \\
\hline 
$A_{12}$    & 0.37    &   0.33 &  0.18   & -0.02        \\
\hline
$A_{21}$    & 0.38     &   0.33 &  0.16   & 5.27$\times 10^{-4}$ \\
\hline
$A_{22}$    & 0.22     &   0.18 &  0.11   & 0.03     \\

\hline \hline 
\end{tabular}
\caption{Asymmetry variable ($A_{ij}$), as defined in eqn.~\ref{asy},
  in the different models for point A and point B. The values quoted
  are at $\sqrt{s}=14 \tev$, after Cut-2.}
\label{tab:asymmetry}
\end{center}
\end{table}

\subsection{Total charge of the four-leptons}

\begin{figure}[h!]
\begin{center}

\centerline{\epsfig{file=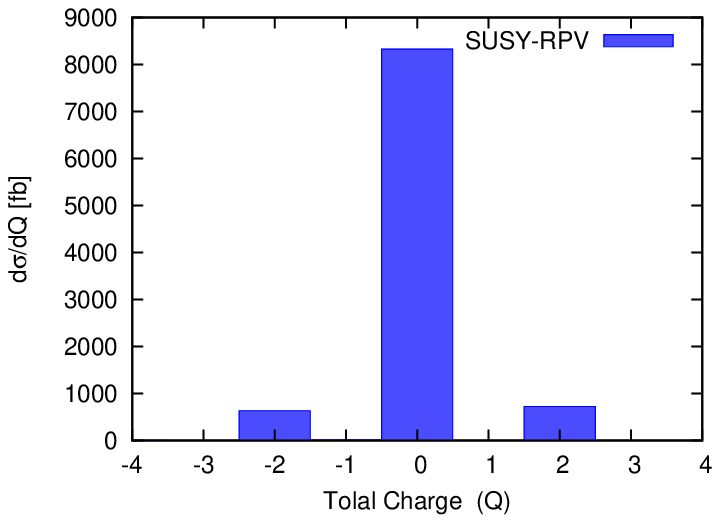,width=6.5cm,height=4.85cm,angle=-0}
\hskip 20pt \epsfig{file=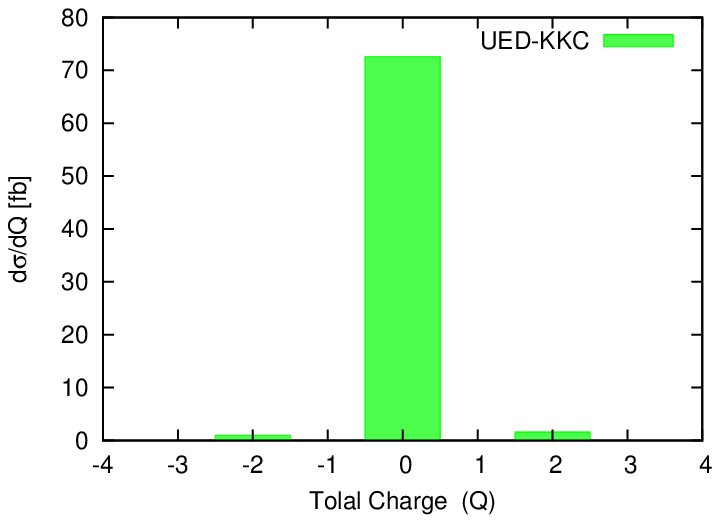,width=6.5cm,height=4.85cm,angle=-0}}

\centerline{\epsfig{file=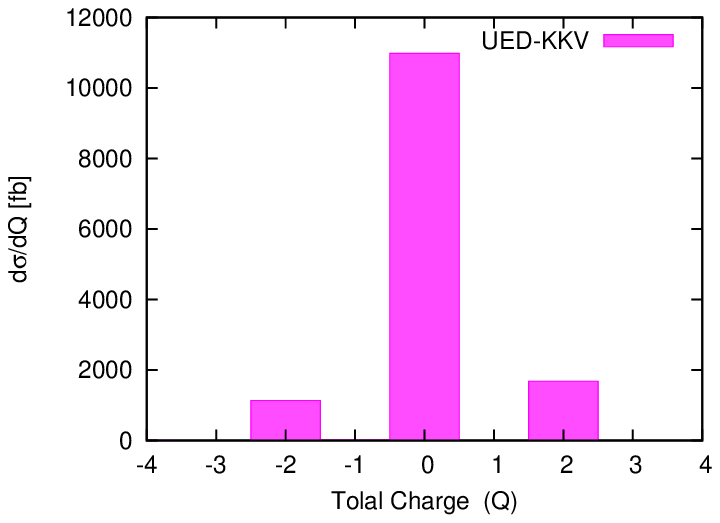,width=6.5cm,height=4.85cm,angle=-0}
\hskip 20pt \epsfig{file=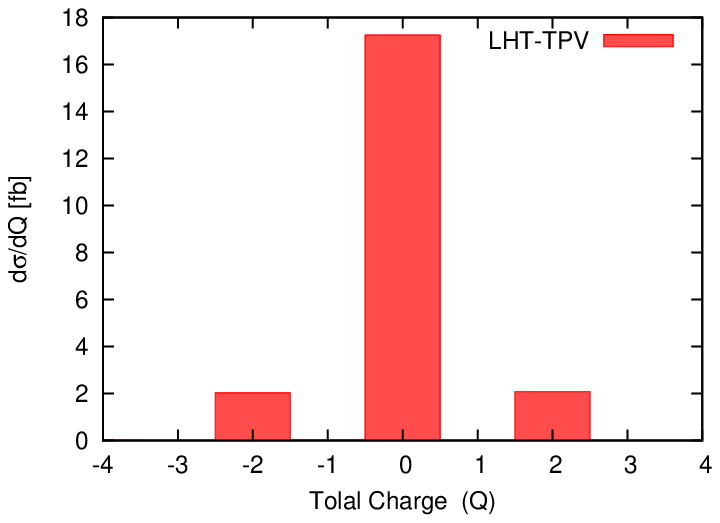,width=6.5cm,height=4.85cm,angle=-0}}
\caption{Total charge distribution of the four-leptons in the
  different models: for point A, at $\sqrt{s}=14 \tev$, after Cut-2.}
\label{fig:Q-600}
\end{center}
\end{figure}

\begin{figure}[h!]
\begin{center}

\centerline{\epsfig{file=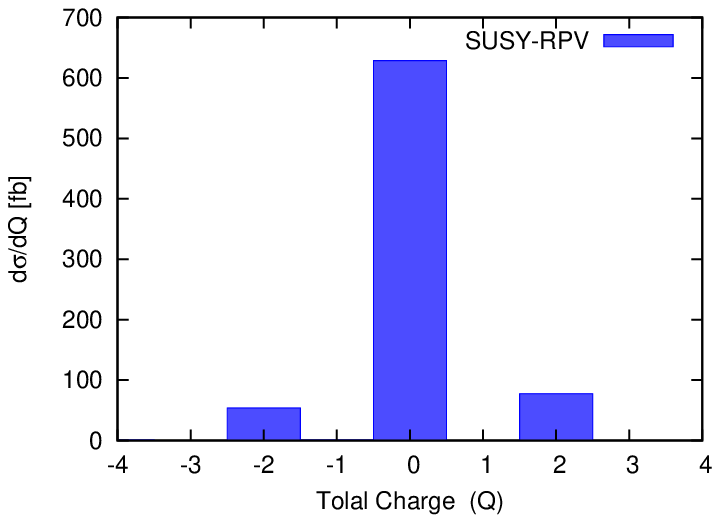,width=6.5cm,height=4.85cm,angle=-0}
\hskip 20pt \epsfig{file=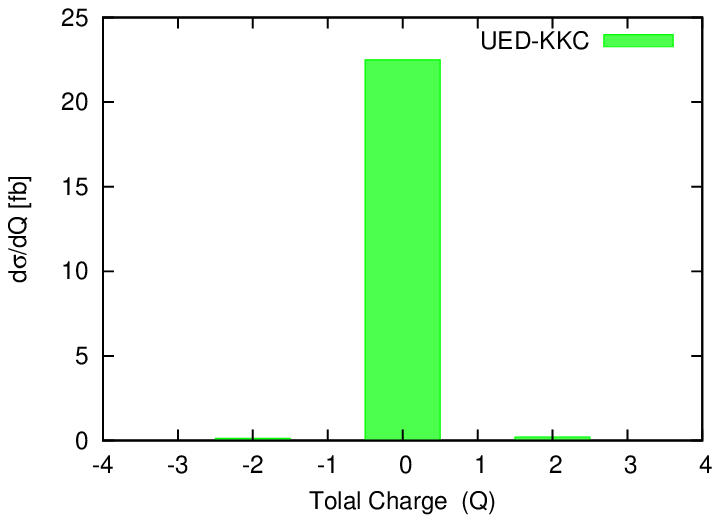,width=6.5cm,height=4.85cm,angle=-0}}

\centerline{\epsfig{file=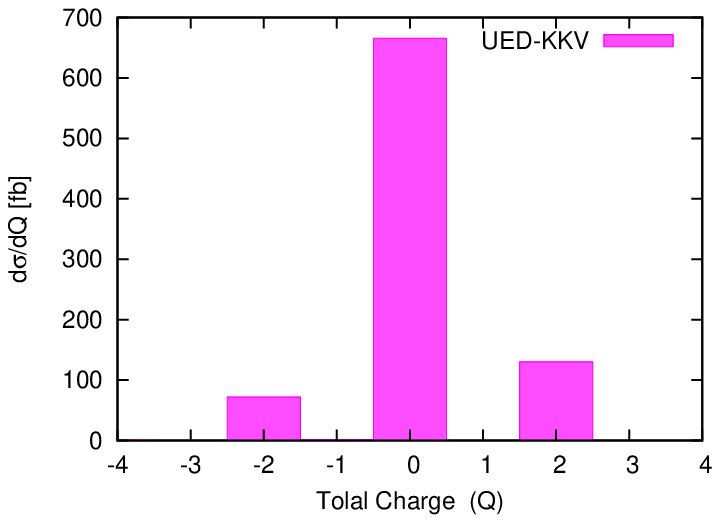,width=6.5cm,height=4.85cm,angle=-0}
\hskip 20pt \epsfig{file=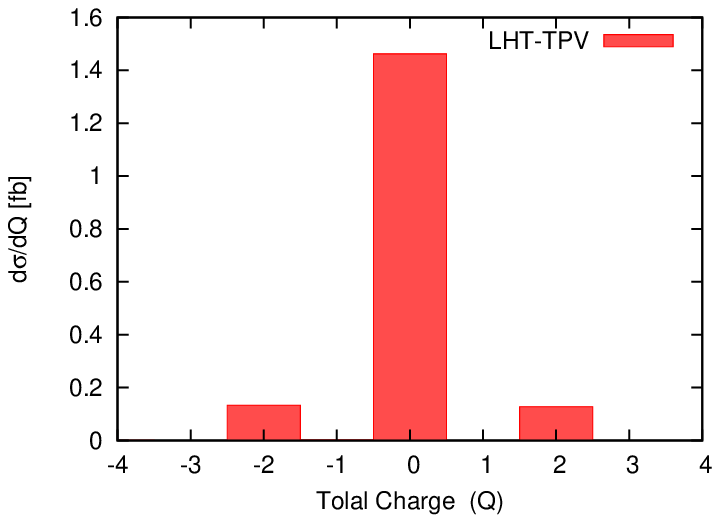,width=6.5cm,height=4.85cm,angle=-0}}
\caption{Same as Figure~\ref{fig:Q-600}, for point B.} 
\label{fig:Q-1000}
\end{center}
\end{figure}
 
Here we consider the variable total charge (Q) of the four leptons
obtained in the general channel $4l+\geq 2j+ \met$. Q can take values
in the set \{-4,-2,0,2,4\}. In
Figures~\ref{fig:Q-600},~\ref{fig:Q-1000} we see the distribution of Q
for the different models. We observe that UED-KKC clearly stands out
as in this case the four leptons come from two cascade decay chains
starting with $Z_1$'s leading to $Q=0$ for all the events. For the RPV
coupling considered for point A, the lightest neutralino in SUSY-RPV
always decays to a pair of leptons and a neutrino. If no other leptons
come from the cascade, these four leptons will have a total charge
$Q=0$. On the other hand, if some additional leptons come from the
decays of charginos or sleptons in the cascade, the lepton
multiplicity in those events will be higher than four. It is however
possible in a certain fraction of events to have a pair of leptons
(coming from the decay of a single boosted neutralino) to be so highly
collimated that they do not pass the lepton-lepton separation
cut. These events might lead to a total charge $Q=\pm2$. For LHT-TPV
different combinations of $W$ and $Z$ decays can lead to four-lepton
events. Hence, the expected values of Q are (-2,0,2). For UED-KKV, the
leptons coming from the decay of $\gamma_1$ are always oppositely
charged. But, it is possible to obtain $W_1$'s of either charge from
the cascades, giving rise to same-sign lepton pairs, over and above
those coming from $\gamma_1$. This leads to $4l$ events with
$Q=\pm2$. Hence, essentially, from the distribution of Q, the UED-KKC
model can be separated from the rest.

\section{Five or higher lepton events}
\label{sec:5l}
\begin{table}[h]

\begin{center}

\begin{tabular}{||c|c|c|c||}
\hline \hline
\multicolumn{4}{||c||}{Point A}\\\hline\hline
Model           & $\sigma(4l)$ in fb & $\sigma(\geq 5l)$ in fb & $\frac{\sigma(\geq 5l)}{\sigma(4l)}$\\\hline\hline
SUSY-RPV & 9450.91 & 3285.85 & 0.35\\
UED-KKC  & 163.36 & 1.57 & 0.01 \\
UED-KKV  & 14008.93 & 4015.64 & 0.29\\
LHT-TPV  & 21.09 & 2.82 & 0.13 \\\hline\hline
\multicolumn{4}{||c||}{Point B}\\\hline\hline
SUSY-RPV & 756.00 & 450.96 &0.60\\
UED-KKC  & 26.24 & 0.12 & 0.005 \\
UED-KKV  & 872.94 & 287.05 & 0.33\\
LHT-TPV  & 1.71 & 0.24 & 0.14 \\\hline\hline
\end{tabular}

\end{center}
\caption{Cross-section of four and five or more lepton events and the
  ratio $\frac{\sigma(\geq 5l)}{\sigma(4l)}$ for the different
  scenarios under consideration. The values tabulated are for point A
  and point B after isolation cuts and $p_T^l>10$ GeV for all leptons,
  at $\sqrt{s}=14 \tev$.}
\label{table:sigma4l_5l}
\end{table}

For all of the models discussed here, additional leptons, over and
above those coming from sources discussed in the previous subsections,
can come from various stages in the cascades. In
Table~\ref{table:sigma4l_5l}, the rates for final states with five or
more leptons are presented for our chosen benchmark points. The $p_T
(E_T)$ cut has been uniformly maintained at 10 GeV for each
lepton. The ratios of the five or higher lepton event rates to those
for four-leptons are also presented.  The SM backgrounds for channels
with lepton multiplicity greater than four are negligible.

The rates for such added profusion of leptons depend on the respective
spectra. The deciding factor here is the leptonic branching ratio for
$Z_2$-odd particles at intermediate stages of the cascade. In this
respect, UED-KKC and LHT-TPV fare worse. For UED-KKC, five-lepton
events are not expected from the production and decays of the $n=1$
KK-particles. Hence, the very small number of events with lepton
multiplicity higher than four come from the decays of B-mesons, pions
or photons. For LHT-TPV, with our choice of parameters, there are few
additional leptons coming from the intermediate steps in the
cascade. Therefore, in most of the $5l$ events found, the leptons are
entirely coming from the decay of the two $A_H$'s produced. One should
note that this rate will increase for other choices of parameters,
especially when the T-odd gauge bosons and the T-odd leptons are
lighter than the T-odd quarks.  For the SUSY-RPV examples considered,
a proliferation of multilepton events is a well-known fact, and that
is reflected in the high value of $\frac{\sigma(\geq
  5l)}{\sigma(4l)}$. For point B, the ratio is higher (0.6) than for
point A (0.35). This is because, as can be seen in
section~\ref{num-results}, in the former case, the SUSY mass spectrum
is such that the chargino can decay to sleptons/sneutrinos, whereas in
the later case such decays are not allowed by phase-space
considerations. This leads to a higher branching fraction of obtaining
a lepton from the cascade, thereby leading to relatively larger rates
for events with higher lepton multiplicity in point B. For, UED-KKV,
the ratio does not change significantly as we change the
mass-scale. Therefore, we conclude that from this ratio, one can
always single out the UED-KKC model. In the other models, there is
enough room to change the ratios by adjusting the mass-spectra
suitably.

\section{Other possibilities}
\label{sec:others}

In addition to the kinds of new particle spectrum considered
above, other possibilities of low missing energy look-alikes 
may offer themselves for detection as well as discrimination
at the LHC. A few illustrative cases are discussed below.

In L-violating SUSY, we have a host of other possibilities. We
considered the case of a neutralino LSP with a $\lambda$-type RPV
coupling in our analysis. In the presence of $\lambda^{\prime}$-type
couplings instead, a neutralino will decay to two quarks and a charged
lepton/ neutrino. Hence, the rate of four-lepton events will
decrease. In addition, we shall not see the asymmetric peaking
behaviour observed in the $cos{\theta_{ij}}$ distributions, too, since
the lepton-pairs in this case will not be coming from the decay of a
single boosted particle. Similarly, the edge in the dilepton invariant
mass distributions will also not be present. There are, however, two
features which distinguish this scenario from the rest. Firstly, the
total charge distribution will now show a significant increase for
$Q=\pm 4$. Secondly, with one lepton coming from each lightest
neutralino during the SUSY cascades in two chains, a pair of same-sign
dileptons occurs frequently at the end of the decay chains.  Since the
gluino, too, is of Majorana nature, being thereby able to produce
charginos of either sign, there will be a surge in same-sign trilepton
events observed~\cite{SS3l}.

A stau LSP scenario with $\lambda$-type couplings will also have
similar implications. Here, stau will decay to a lepton and a
neutrino. If we have a stau LSP with $\lambda^{\prime}$-type couplings
instead, the 4l-signal will be suppressed, as the stau will decay to
two quarks in this case. We also note that in presence of relatively
larger values of $\lambda^\prime$-type couplings, there is a region of
the mSUGRA parameter space where the sneutrino can also become an
LSP~\cite{Allanach,Dreiner:sneutrino}. However, the sneutrino in this
case will decay to a pair of quarks thus reducing the possibility of
having four leptons in the final state. Finally, we also remark that
in the presence of bilinear R-parity violating terms, if we choose the
RPV parameters to be consistent with the observed neutrino mass and
mixing patterns, the neutralino LSP can decay to $W\mu$, $W\tau$ or
$Z\nu$~\cite{R_M_2}. Hence, four-lepton events will again be viable
with a peak at $M_Z$ in the OS dilepton invariant mass
distributions. But there will not be any peak in the corresponding
$M_{4l}$ distribution. This last feature thus seems to be generically
true in SUSY-RPV.

In the framework of a minimal UED model with KK-parity conservation,
$\gamma_1$ is the LKP and characteristic signatures of this scenario
were discussed throughout this article. The mUED model is based on the
simplifying assumption that the boundary kinetic terms vanish at the
cut-off scale $\Lambda$. This assumption restricts the mUED spectrum
in such a way that the splittings among different KK excitations are
small enough to result in soft final state particles and thus low
missing $E_T$. As discussed in section~\ref{UED-KKC}, the boundary
terms receive divergent contributions and thus require
counterterms. The finite parts of these counterterms remain
undetermined and are therefore free parameters of the theory. This
additional freedom could be exploited to end up with an unconstrained
UED (UUED) scenario with several different possibilities:

\begin{itemize}
\item Particular combinations of the aforementioned free parameters
  can remove the degeneracy of the mUED mass spectrum. This results in
  harder leptons and jets in the final state and thus gives rise to a
  harder missing-$E_T$ distribution. Such a version of the UED model
  no longer falls in the category of {\em low missing energy
    look-alikes}.  The corresponding criteria for large missing-$E_T$
  scenarios will have to be applied in this situation
  ~\cite{Datta_Kane}.

\item The values of the additional parameters could be chosen in such
  a way that even though the mass spectrum remains degenerate as in
  the case of mUED, the mass of $\nu_1$ becomes smaller than
  $M_{\gamma_1}$. This leads to a scenario with $\nu_1$ as the LKP and
  therefore, a possible candidate for cold dark matter. In this case,
  the $\gamma_1$'s produced at the end of decay cascades will further
  decay into $\nu \nu_1$ pairs, resulting in similar phenomenology as
  in the case of mUED explored in this work. However, the four-lepton
  event rate in this case will be significantly reduced due to the
  fact that $Z_1$ mostly decays to $\nu \nu_1$ pairs, as this channel
  will now have enhanced phase-space available as compared to $lL_1$.

\item One can also have such a combination of parameters that only the
  positions of $Z_1$ and $W_1^\pm$ in the $n=1$ mass spectrum are
  altered. The first possibility is that $Z_1$ and $W_1^\pm$ are
  heavier than the $n=1$ quarks. In such situations, $n=1$ quarks will
  decay into $qlL_1$ via tree-level 3-body modes. $L_1$ subsequently
  decays into $l\gamma_1$ and combining the two cascade decay chains,
  one can then obtain four-lepton signals. However, the opposite-sign
  dilepton invariant mass distributions will not show any invariant
  mass edge in this scenario. The other possibility is that $Z_1$ and
  $W_1^\pm$ may become lighter than the $n=1$ leptons. Here, the
  possible decay modes of interest are $Z_1\rightarrow ll\gamma_1$ and
  $W_1^\pm \rightarrow l\nu\gamma_1$. Phenomenologically, this
  scenario will again be quite similar to the mUED cases we have
  studied.
\end{itemize}

In the framework of UED-KKV, the masses of the KK-fermions have a
small dependence on the value of the KK-parity violating coupling
$h$. As shown in Ref.~\cite{Bhattacherjee}, as we increase the value
of $h$, the KK-fermion masses decrease. As the $n=1$ quarks are much
heavier than $Z_1$ and $W^\pm_1$ in mUED with KK-parity conserved, the
presence of non-zero KK-violating coupling $h$ does not lead to any
appreciable change in the excited quark masses vis-a-vis those of the
remaining particles. Therefore, the decay patterns of the $n=1$ quarks
remain the same. However, for smaller $R^{-1}$, singlet $n=1$ leptons
and the $\gamma_1$ are almost degenerate. Therefore, in addition to
having a smaller $R^{-1}$, if we have a larger value of $h$, singlet
$n=1$ leptons may become lighter than $\gamma_1$. For example, this
can happen for $R^{-1}=500 \gev$ and $h>0.01$. Here, the only possible
decay channel available for these singlet $n=1$ leptons is via the
KK-parity violating mode to a SM lepton of the same flavour and a
$Z/\gamma^*$. The phenomenology of this scenario will be significantly
different from the one we have considered so far with smaller values
of $h$ ($h\sim 0.001$) and a $\gamma_1$ LKP, and is therefore an
interesting possibility for further studies.

In LHT-TPV, as we have two free parameters $\kappa_q$ and $\kappa_l$
which determine the masses of the T-odd quarks and leptons, the
possibility of obtaining a fermionic LTP exists. In such cases, via
the T-parity breaking terms, these fermions might decay to standard
model particles. As observed in Ref.~\cite{Freitas_Schwaller_Wyler},
there are two possible ways in which these fermionic LTP's can
decay. As the WZW term breaks T-parity only in the gauge sector
directly, a T-odd lightest fermion can decay by a three-body mode
mediated by a virtual T-odd gauge boson. There is also the possibility
of having loop-induced two-body decay modes. Therefore, the major
decay channels are $f_H\rightarrow Zf$, $f_H\rightarrow W \tilde{f}$
and $f_H\rightarrow A^{*}_Hf$. In the last case, the off-shell $A_H$
might again decay to four-leptons via $ZZ$, but there will not be any
peak in the $M_{4l}$ distribution. On the other hand, if the LTP is a
T-odd charged lepton, one can observe a spectacular five-lepton
peak. The other possible decay modes will also lead to $4l$ and $5l$
events in varied rates, although the quantitative predictions of these
will require us to determine the relative importance of the three-body
and the loop-induced two body decay modes of $f_H$. A detailed
phenomenology of this fermionic LTP scenario will be studied in a
future work.

\section{Summary and conclusions}
\label{summary}

We have considered four characteristic scenarios which can pass off as
low missing energy look-alikes at the LHC. After convincing the reader
that UED-KKC in its minimal form, in spite of containing a stable
invisible particle, often falls in this category, we have studied the
contribution of each scenario to events with four or more
leptons. Since total rates alone, at the four-lepton level at least,
can mislead one in the process of discrimination, we have resorted to
kinematic features of final states in the different cases.

The first of these is the four-lepton invariant mass distribution.  On
this, we have found that the models get divided into two categories,
depending on whether the four leptons show an invariant mass peak or
not. While LHT-TPV and UED-KKV are in the first category, SUSY-RPV and
and UED-KKC belong to the second one, thus offering a clear
distinction.

For distinguishing between models within each category, we have used
three observable quantities, namely, angular correlation and invariant
mass of opposite-charge lepton pairs, and the total charge of the four
observed leptons. It is found that angular correlations as well as the
total charge causes SUSY-RPV to stand out quite clearly with respect
to minimal UED-KKC. UED-KKV, too, stands out distinctly from the
others, as far as angular correlations are concerned. This, however,
still leaves room for discrimination. This happens, for example, where
LHT-TPV has such a spectrum that the heavy photon LTP does not decay
dominantly into four leptons. The distinction of this scenario with
SUSY-RPV and UED-KKC is still possible using the pairwise invariant
mass distribution of oppositely charged dileptons. We have also found
that the relative positions of the dilepton vs. four-lepton invariant
mass peaks, and the existence of an `edge' in the dilepton invariant
mass distribution lead to useful discriminating criteria. In addition,
the ratio of five or higher lepton event rates to those for four
leptons sets UED-KKC apart from the other scenarios.

As mentioned before, we haven't paid particular attention so far to the flavour content of the four leptons. In SUSY-RPV, depending upon the L-violating coupling chosen, the flavour content will change. However one can discover a  pattern in the fraction of events with $4e$, $4 \mu$, $2e2\mu$, $3e1\mu$ or $1e3\mu$, for specific RPV couplings (here $e$ stands for the electron or positron and $\mu$ stands for the muon or anti-muon, and, for example, $3e1\mu$ means a four-lepton event with $3$ $e$'s and $1$ $\mu$). For the $\lambda_{122}$ coupling that we chose for our analysis, we obtain, as expected, around $47.5 \%$ events to be of $1e3\mu$-type, whereas $2e2\mu$ and $4\mu$ being $\sim 25\%$ each. No events are expected to be of $3e1\mu$ or $4e$-type. This is certainly a notable feature, but as observed above, these flavour ratios will change if we change the L-violating coupling. For UED-KKC one will always obtain $4e$, $4\mu$ or $2e2\mu$-type events, while for UED-KKV all flavour combinations are possible. For LHT-TPV, depending upon the point of parameter space one is in, all flavour combinations are once again possible with varying fractions. Thus, although flavour content of the leptons can sometimes be useful, they might not be a very robust feature of the models, except for the case of UED-KKC. 

We have also made a set of qualitative observations on other related
scenarios such as RPV via $\lambda^\prime$-type or bilinear terms,
situations with stau or sneutrino LSP, UED-KKC with an unconstrained
particle spectrum and different LTP (LKP) in LHT-TPV (UED-KKV). The
qualitative changes that some of these scenarios entailed have been
pointed out. On the whole, while the various theoretical models in
their `minimal' forms offer clear methods for distinction, a confusion
can always be created in variants of the models with various degrees
of complications.  The same observation holds also for theories
predicting large missing $E_T$.  Thus one may perhaps conclude that,
while some striking qualitative differences await one in the approach
from `data to minimal theories', there is no alternative to a
threadbare analysis of the mass spectrum, possibly linking spin
information alongside, if one really has to exhaust all possibilities
that nature may have in store. Furthermore, the availability of data
with high statistics, enabled by large accumulated luminosity, is of
great importance.  All this is likely to keep the highly challenging
character of the LHC experiment alive till the last day of its run.

\appendix
 \renewcommand{\theequation}{A-\arabic{equation}}
  \setcounter{equation}{0}  

\section*{Appendix}
\section{Description of the various models considered}
\subsection{SUSY with R-parity violation (SUSY-RPV)}
The MSSM superpotential is given by \cite{S.P.Martin1}
\be 
W_{MSSM} = {y^l}_{ij} L_i H_1 {\bar {E_j}} + {y^d}_{ij} Q_i H_1 {\bar {D_j}}+ {y^u}_{ij} Q_i H_2 {\bar {U_j}}+\mu H_1 H_2 
\label{eqn:mssm-sup}
\ee where $H_1$ and $H_2$ are the two Higgs superfields, L and Q are
the SU(2)-doublet lepton and quark superfields and E, U, and D are the
singlet lepton, up-type quark, and down-type quark superfields,
respectively. $\mu$ is the Higgsino mass parameter and $y_{ij}$'s are
the strengths of the Yukawa interactions.

If lepton and baryon number are allowed to be broken, the above
superpotential can be augmented by the following terms~\cite{Barbier}:
\be \label{RPV} W_{\Rp} = \lambda_{ijk}L_iL_j{\bar{E}}_k +
    {\lambda^{\prime}_{ijk}}L_iQ_j{\bar D}_k+ {\lambda^{\prime
        \prime}}_{ijk}{\bar U}_i{\bar D}_j{\bar D}_k+ {\epsilon}_i L_i
    H_2. \label{supp} \ee
where i, j, and k are flavour indices. Here the ${\lambda^{\prime
    \prime}}_{ijk}$ term leads to baryon number (B) violating
interactions while the other three terms lead to the violation of
lepton number (L). In MSSM, one imposes an additional discrete
multiplicative symmetry known as R-parity which prevents any such term
in the superpotential. R-parity is defined as $R_p=(-1)^{3B+L+2S}$,
where $S$ is the spin of the corresponding particle. This prevents,
for example, terms which can lead to fast proton decay (although there
are dangerous ``R-even'' dimension-five operators which can still lead
to the decay of proton). However, the purpose is equally well-served
if only {\it one} between B and L is conserved, R-parity violating
(RPV) SUSY models are constructed with this in view.  Of these, the
versions containing L-violating interactions, trilinear and/or
bilinear, have the added motivation of offering explanations of
neutrino masses and mixing.

The consequences of RPV interactions have been explored extensively in
the literature~\cite{Barbier}. Especially, when L-violation takes
place, although the conventional large missing $E_T$ signature of SUSY
is degraded, the possibility of obtaining multilepton final states is
enhanced~\cite{Baer}. Recently, it has also been demonstrated that in
presence of these L-violating operators, rather striking same-sign
three-and four-lepton final states, which are free from SM
backgrounds, are expected in large rates at the LHC ~\cite{SS3l}.

In presence of the $\lambda$-type couplings, if the neutralino is the
lightest supersymmetric particle (LSP), it will decay to a pair of
leptons and a neutrino. Thus, starting from the pair production of
squarks/gluinos in the initial parton level $2\rightarrow2$ hard
scattering, we can obtain two neutralinos at the end of the decay
cascade, which in turn can give rise to a four-lepton final state with
unsuppressed rate for every SUSY process.  Also, at least two
additional jets will always be present from the decay of the squark
pair. Thus, one can very easily obtain the $4l+nj+\met$ ($n\geq2$). If
in addition to this, another lepton is produced from the cascade decay
of a chargino, we can also obtain a five-lepton final state.

The $\lambda^{\prime}$-type interactions, on the other hand, cause a
neutralino LSP to decay into a lepton and two parton-level jets, this
giving a dilepton final state for every SUSY particle production
process.  Four-lepton final states can arise through two additional
leptons produced in cascade, after the initial $2\rightarrow2$
process.  Understandably, the probability of obtaining a four-lepton
final state is suppressed compared to dilepton and trileptons.

The bilinear terms $\epsilon_i L_i H_2$ imply mixing between neutrinos
and neutralinos as also between charged leptons and
charginos~\cite{R_M_2}. Consequently, the lightest neutralino LSP can
decay into a neutrino and the Z or a charged lepton and the W, the
latter mode being of larger branching ratio. Thus, modulo the leptonic
branching ratios of W-and Z-decay, one can have four-lepton final
states, with an additional lepton in the cascade to account for the
somewhat suppressed five-lepton final states in such a scenario.

It is clear from the above discussion that four-and five-lepton states
are expected to have the highest rates in a situation with R-parity is
broken through $\lambda$-type interactions alone, with two indices in
$\lambda_{ijk}$ being either 1 or 2. We, therefore, use this scenario
as the benchmark for distinction from other theories through
multilepton events.  In situations including the other L-violating
terms, the dilution in the branching ratio will affect the total rates
of such final states (and, as we shall discuss later, the total rates
are not very good guidelines in model distinction anyway). However,
the kinematical observables on which our proposed distinction criteria
are based are likely to remain mostly quite similar.  We will comment
on some special cases, including those where the lightest neutralino
is not the LSP, in section~\ref{sec:others}.

\subsection{Minimal universal extra dimension with KK-parity conservation (UED-KKC)}
\label{UED-KKC}
A rather exciting development in physics beyond the Standard Model
(SM) in the last few years is the formulation of theories with compact
spacelike extra dimensions whose phenomenology can be tested at the
TeV scale. The idea is based on concepts first introduced by Kaluza
and Klein \cite{ED:kaluza} in the 1920's.  Extra-dimensional theories
can be divided into two main classes. The first includes those where
the Standard Model (SM) fields are confined to a (3 + 1) dimensional
subspace (3-brane) of the full manifold. Models with the extra
spacelike dimensions being both flat \cite{add} or warped \cite{rs}
fall in this category, although there have been numerous attempts to
put some or all of the fields in the `bulk' even within the ambits of
these theories. On the other hand, there is a class of models known as
Universal Extra Dimension(s) (UED) \cite{acd}, where all of the SM
fields can access the additional dimensions.

In the minimal version of UED (mUED), there is only one extra
dimension $y$ compactified on a circle of radius $R$ ($S_1$ symmetry).
The need to introduce chiral fermions in the resulting
four-dimensional effective theory prompts one to impose additional
conditions on the extra dimension. This condition is known as
`orbifolding' where two diametrically opposite ends of the compact
dimension are connected by an axis about which there is a reflection
symmetry.

The $Z_2$ symmetry breaks the translational invariance along the fifth
dimension (denoted by the co-ordinate y) and generates two fixed
points at $y=0$ and $y=\pi R$.  From a four-dimensional viewpoint,
every field will then have an infinite tower of Kaluza-Klein (KK)
modes, the zero modes being identified as the corresponding SM
states. The spectrum is essentially governed by $R^{-1}$ where R is
the radius of the extra dimension.

UED scenarios can have a number of interesting phenomenological
implications. These include a new mechanism of supersymmetry breaking
\cite{antoniadis1}, relaxation of the upper limit of the lightest
supersymmetric neutral Higgs \cite{relax}, addressing the issue of
fermion mass hierarchy from a different perspective
\cite{Arkani-Hamed:1999dc} and lowering the unification scale down to
a few TeVs \cite{dienes,dienes2,blitz}.

In the absence of the orbifold fixed points, the component of momentum
along the extra direction is conserved. From a four-dimensional
perspective, this implies KK number conservation, where KK number is
given by the position of an excited state in the tower. However, the
presence of the two orbifold fixed points breaks the translational
symmetry along the compact dimension, and KK number is consequently
violated.  In principle, one may have some additional interaction
terms localised at these fixed points, causing mixing among different
KK states. However, if these interactions are symmetric under the
exchange of the fixed points (this is another $Z_2$ symmetry, not to
be confused with the $Z_2$ of $y \leftrightarrow$ $-y$), the
conservation of KK number breaks down to the conservation of KK
parity\footnote {In principle, it is possible to have fixed point
  localized operators that are asymmetric in nature~\cite{acd}. This
  could violate KK-parity, analogous to the R-parity violation in
  supersymmetry. In absence of KK-parity the phenomenology of UED will
  be drastically different and will be discussed in brief in the next
  subsection.}, given by $(- 1)^n$, where $n$ is the KK number.  This
not only implies that level-one KK-modes, the lightest among the new
particles, are always produced in pairs, but also ensures that the KK
modes do not affect electroweak processes at the tree level. The
multiplicatively conserved nature of KK-parity implies that the
lightest among the first excitations of the SM fields is stable, and a
potential dark matter candidate~\cite{Servant:2002aq}.

The tree-level mass of a level-$n$ KK particle is given by $m_n^2 =
m_0^2 + {n^2}/{R^2}$, where $m_0$ is the mass associated with the
corresponding SM field. Therefore, the tree level mUED spectrum is
extremely degenerate and, to start with, the first excitation of any
massless SM particle can be the lightest KK-odd particle (LKP). In
practice, radiative corrections \cite{Cheng:2002iz} play an important
role in determining the actual spectrum. The correction term can be
finite (bulk correction) or it may depend on $\Lambda$, the cut-off
scale of the model (boundary correction). Bulk corrections arise due
to the winding of the internal lines in a loop around the compactified
direction~\cite{Cheng:2002iz}, and are nonzero and finite only for the
gauge boson KK-excitations. On the other hand, the boundary
corrections are not finite, but are logarithmically divergent.  They
are just the counterterms of the total orbifold correction, with the
finite parts being completely unknown. Assuming that the boundary
kinetic terms vanish at the cutoff scale $\Lambda$ the corrections
from the boundary terms, for a renormalization scale $\mu$, are
proportional to $L_0 \equiv \ln (\Lambda^2 /\mu^2) $.  The bulk and
boundary corrections for level-n doublet quarks and leptons ($Q_n$ and
$L_n$), singlet quarks and leptons ($q_n$ and $e_n$) and KK gauge
bosons ($g_n$, $W_n$, $Z_n$ and $B_n$) are given by,
\begin{itemize}
\item Bulk corrections:
\begin{eqnarray}
\delta\, (m_{B_n}^2) &=&  -\frac{39 \,\zeta(3) \, a_1}{2 \, \pi^2 \, R^2} \ ,\nonumber\\
\delta\, (m_{W_n}^2) &=&    -\frac{5\,\zeta(3) \, a_2}{2 \, \pi^2 \, R^2}\ ,\nonumber
\\
\delta\, (m_{g_n}^2) &=&   -\frac{3\,\zeta(3) \, a_3}{2 \, \pi^2 \, R^2}\ ,
\label{delta0}
\end{eqnarray}
where, $\zeta(3)=\sum_{n=1}^{\infty}{1}/{n^3}\simeq 1.202$.

\item Boundary corrections:
\begin{eqnarray}
\bar{\delta}\, m_{Q_n} &=&  
m_n \,\left(3\, a_3
+ \frac{27}{16}\,a_2 + \frac{a_1}{16}
\right) \,L_0 \ ,~~ \bar{\delta}\, m_{u_n}~ =~ m_n \,\left(3\, a_3
  + a_1
\right) \, L_0 \ , \nonumber\\
\bar{\delta}\, m_{d_n} &=&  m_n \,\left(3\, a_3
  + \frac{a_1}{4}
\right) \, L_0 \ ,~~~~~~~~~~~~ 
\bar{\delta}\, m_{L_n}~ = ~ m_n \,\left(
 \frac{27}{16}\,a_2 + \frac{9}{16}\,a_1
\right) \, L_0 \ , \nonumber\\
\bar{\delta}\, m_{e_n} &=& 
 \frac{9\,a_1}{4}
\,  m_n \,L_0 \ ,~~~~~~~~~~~~~~~~~~~~~
\bar{\delta}\, (m_{B_n}^2)~ =~ -\frac{a_1}{6}\, m_n^2\,  L_0 \nonumber
 \\
\bar{\delta}\, (m_{W_n}^2) &=& \frac{15\, a_2}{2}
\, m_n^2\, L_0 \ , ~~~~~~~~~~~~~~~~~~~~
 \bar{\delta}\, (m_{g_n}^2)~ =~ \frac{23\, a_3}{2} \,  m_n^2 \,
L_0 \ , \nonumber\\
\bar{\delta}\, (m_{H_n}^2) &=& m_n^2 \, \left(\frac{3}{2}\, a_2   
+ \frac{3}{4}\, a_1 - \frac{\lambda_H}{16\pi^2} \right)
L_0 + \overline{m}_H^2\ ,
\end{eqnarray}
\end{itemize}
where $a_i \equiv g_i^2 / 16 \, \pi^2 \ , i = 1\dots 3$, $g_i$ being
the respective gauge coupling constants and $\overline{m}_H^2$ is the
boundary term for the Higgs scalar, which has been chosen to be zero
in our study.

These radiative corrections partially remove the degeneracy in the
spectrum \cite{Cheng:2002iz} and, over most of the parameter space,
$\gamma_1$, the first excitation of the hypercharge gauge boson ($B$),
is the LKP
\footnote {The KK Weinberg angle is small so that $B_1$ $\approx$
  $\gamma_1$ and $W^{3}_1$ $\approx$ $Z_1$. }.  The $\gamma_1$ can
produce the right amount of relic density and turns out to be a good
dark matter candidate \cite{Servant:2002aq}. The mass of $\gamma_1$ is
approximately $R^{-1}$ and hence the overclosure of the universe puts
an upper bound on $R^{-1} < $ 1400 GeV. The lower limit on $R^{-1}$
comes from the low energy observables and direct search of new
particles at the Tevatron. Constraints from $g-2$ of the muon
\cite{nath}, flavour changing neutral currents \cite{chk,buras,desh},
$Z \to b\bar{b}$ \cite{santa}, the $\rho$ parameter
\cite{acd,appel-yee}, other electroweak precision tests \cite{ewued},
etc. imply that $R^{-1}~\gtap~300$ GeV. The masses of KK particles are
also dependent on $\Lambda$, the cut-off of UED as an effective
theory, which is essentially a free parameter. One loop corrected
$SU(3)$, $SU(2)$ and $U(1)$ gauge couplings show power law running in
the mUED model and almost meet at the scale $\Lambda$= $20 R^{-1}$
\cite{dienes}. Thus one often takes $\Lambda =20 R^{-1}$ as the
cut-off of the model. If one does not demand such unification, one can
extend the value of $\Lambda$ to about $40 R^{-1}$, above which the
$U(1)$ coupling becomes nonperturbative.

After incorporating the radiative effects, the typical UED spectrum
shows that the coloured KK states are on top of the spectrum. Among
them, $g_1$, the $n = 1$ gluon, is the heaviest. It can decay to both
$n=1$ singlet ($q_1$) and doublet ($Q_1$) quarks with almost the same
branching ratio, although there is a slight kinematic preference for
the singlet channel. $q_1$ can decay only to $\gamma_1$ and an SM
quark.  On the other hand, a doublet quark $Q_1$ decays mostly to
$W_1$ or $Z_1$. Hadronic decay modes of $W_1$ and $Z_1$ are closed
kinematically, so that they decay to different $n=1$ doublet leptons.
Similarly $Z_1$ can decay only to $L_1 l$ or $\nu_1 {\nu}$. The KK
leptons finally decay to the $\gamma_1$ and SM leptons. Thus, the
principal signals of such a scenario are $n$ jets + $m$ leptons + MET,
where $m$ can vary from 1 to 4.

\subsection{mUED with KK-parity violation (UED-KKV)}
\label{UED-KKV}
Let us now briefly introduce KK-parity violation and its
phenomenological consequences.  As discussed in the previous
subsection, operators localised at the orbifold fixed points give rise
to mixing between different KK-states. In presence of such operators
that are symmetric under exchange of the fixed points, even and odd
KK-parity states mix separately, so that KK-parity is still a
conserved quantity.  However, it is possible to have fixed point
operators which are asymmetric in nature, leading to the violation of
KK-parity \cite{acd}. The phenomenology of KK-parity violation was
studied in some detail in Ref. \cite{Bhattacherjee}. KK-parity
violating couplings allow the LKP to decay into SM particles.

The asymmetry in the localised operators can be introduced by adding
the following extra contribution to the operators localized at the
point $y=0$:

\begin{equation}
{\cal L}_f=\frac{\lambda}{\Lambda}\int \left[i\bar \psi \Gamma^\alpha
D_\alpha
\psi-i(D_\alpha\bar\psi)\Gamma^\alpha\psi\right]\delta(y)dy,
\label{eq:kkpv}
\end{equation} 
where, $\psi(x^\mu,y)$ is the 5-dimensional fermion field,
$\Gamma^\alpha$ are the gamma matrices and $D_\alpha$ are the
covariant derivatives defined in (4+1) dimensions, $\lambda$ is the
strength of the KK-parity violating coupling, and $\Lambda$ is the
cut-off scale. The above operator has the following consequences:

\begin{itemize}
\item Equation \ref{eq:kkpv} gives rise to KK-parity violating mixing
  between KK-even and odd states. Of them, the mixing between $n=0$
  and $n=1$ states are most relevant from the angle of the LHC
  phenomenology. The admixture of $n\ge2$ KK-states with the $n=0$
  state is suppressed by the higher masses of the former.

\item Equation \ref{eq:kkpv} leads to the coupling between two
  KK-fermions ($\psi^{(l)}$ and $\psi^{(m)}$) and a gauge boson
  ($A_{\mu}^{(n)}$): $ihg \bar\psi^{(l)}\gamma^\mu
  A_{\mu}^{(n)}\psi^{(m)}$, where $h$ is a dimensionless coupling
  parametrised as $h=\lambda/(2\pi \Lambda R)$ and $g$ is the gauge
  coupling. This coupling can be KK-parity violating, if $(-1)^{n+m+l}
  = -1$.

\item In the KK-basis, interaction terms involving one $n=0$ and two
  $n=1$ particles are allowed by KK-parity. As an example consider the
  terms $ig \bar\psi^{(1)}_{KK}\gamma^\mu
  A_{\mu}^{(1)}\psi^{(0)}_{KK}$, $ig \bar\psi^{(1)}_{KK}\gamma^\mu
  A_{\mu}^{(0)}\psi^{(1)}_{KK}$ etc., where $\psi^{(n)}_{KK}$ is
  level-$n$ fermion in the KK-basis. As a result of the mixing between
  $n=0~{\rm and}~1$ fermions , the above interactions give rise to
  KK-parity violating couplings between two $n=0$ particles with one
  $n=1$ particle in the mass eigenstate basis.

\end{itemize}

We have only considered the consequences of the fermionic kinetic
terms localized at the point $y=0$. However, in principle, one could
also introduce kinetic terms for the 5-dimensional gauge bosons at
$y=0$. Such terms would lead to KK-parity violating decay of $n=1$
gauge bosons into a pair of SM electroweak gauge bosons. For
$W^{\pm}_1$ and $Z_1$-boson, the KK-parity conserving decays are
overwhelmingly dominant. Therefore, the KK-parity violating decays of
$W^{\pm}_1$ and $Z_1$-boson are phenomenologically insignificant. On
the other hand, the LKP ($\gamma_1$), being completely $B_1$
dominated, can not have any coupling with the SM $ZZ$ or $W^+W^-$
pairs. Therefore, from the point of view of collider phenomenology,
the KK-parity violating kinetic terms for the 5-dimensional gauge
bosons are not significant.

We have already noted that, as a consequence of KK-parity violation,
the lightest $n=1$ particle decays into two or more SM
particles. Thus, $\gamma_1$ can decay into SM fermion-antifermion
pairs. The KK-parity violating couplings of $\gamma_1$ with the SM
fermions are given by
\begin{equation}
{\cal L}_{KKV}= \frac{eh}{\sqrt 2 {\rm cos}\theta_W}
\bar f [Y_L\gamma_\mu(1-\gamma_5)-Y_R\gamma_\mu(1+\gamma_5)]\gamma^\mu_1 f, 
\end{equation}

where $f$ is a SM fermion and $Y_L~{\rm and }~ Y_R$ are the
hypercharges of $f_L ~{\rm and} f_R$ respectively. For $\lambda \sim
1$ and $\Lambda R \sim 20$, the maximum value of the parameter $h$ is
$h_{max}\sim ~ 0.01$. In our analysis, we take $h$ to be same for all
fermionic flavours. With this assumption, the decay branching fraction
of $\gamma_1$ becomes independent of the value of $h$. We obtain
53.3\% branching fraction for $\gamma_1\to q\bar q$ and 38.9 (7.8)\%
branching fraction for $\gamma_1 \to l \bar l ~(\nu \bar \nu)$. For
small values of the parameter h, the decay branching fractions of
other $n=1$ particles are insensitive to KK-parity violation, since
such effects are suppressed by the cut-off scale $\Lambda$, so that
they cannot compete with the KK-parity conserving modes. Therefore,
the only phenomenological difference between UED-KKC and UED-KKV is
the decay of the LKP and its decay gives rise to additional jets and
leptons in the final states of UED cascades.  Thus the additional
leptons lead to rather clean signatures with the possibility of
reconstructing the $\gamma_1$ as an invariant mass peak. On the whole,
depending on whether either or both of the pair-produced $\gamma_1$'s
decay leptonically, the lepton multiplicity of the ensuing final
states may vary between 0 and 8.

\subsection{Littlest Higgs model with T-parity violation (LHT-TPV)}

Little Higgs models~\cite{LH_original, LH_reviews} have been proposed
a few years ago to explain electroweak symmetry breaking and, in
particular, to solve the so-called little hierarchy
problem~\cite{LEP_paradox}. Among the different versions of this
approach, the littlest Higgs model~\cite{Littlest_Higgs} achieves the
cancellation of quadratic divergences with a minimal number of new
degrees of freedom.  In the LHT, a global $SU(5)$ symmetry is
spontaneously broken down to $SO(5)$ at a scale $f \sim 1\tev$. An
$[SU(2) \times U(1)]^2$ gauge symmetry is imposed, which is
simultaneously broken to the diagonal subgroup $SU(2)_L \times
U(1)_Y$, the latter being identified with the SM gauge group. This
leads to four heavy gauge bosons $W_H^\pm, Z_H$ and $A_H$ with masses
$\sim f$ in addition to the SM gauge fields. The SM Higgs doublet is
part of an assortment of pseudo-Goldstone bosons which result from
breakdown of the global symmetry. This symmetry protects the Higgs
mass from quadratically divergent corrections. The multiplet of
Goldstone bosons contains a heavy $SU(2)$ triplet scalar $\Phi$ as
well. In contrast to SUSY, the new states which cancel the
quadratically divergent contributions to the Higgs mass due to the top
quark, gauge boson and Higgs boson loops, respectively, are heavy
fermions, additional gauge bosons and triplet Higgs states.

In order to comply with strong constraints from electroweak precision
data on the Littlest Higgs model~\cite{LH_EW_tests}, and at the same
time ensure that the scale of new physics be not so high as to cause
the so-called little hierarchy problem, one further imposes a discrete
symmetry called T-parity~\cite{T_parity}. It maps the two pairs of
gauge groups $SU(2)_i \times U(1)_i, i=1,2$ into each other, forcing
the corresponding gauge couplings to be equal, with $g_1 = g_2$ and
$g_1^\prime = g_2^\prime$. All SM particles, including the Higgs
doublet, are even under T-parity, whereas the four additional heavy
gauge bosons and the Higgs triplet are T-odd. The top quark has two
heavy fermionic partners, $T_{+}$ (T-even) and $T_{-}$ (T-odd). For
consistency of the model, one has to introduce the additional heavy,
T-odd vector-like fermions $u^i_H, d^i_H, e^i_H$ and $\nu^i_H$
($i=1,2,3$) for each SM quark and lepton field. For further details on
the LHT, we refer the reader to
references~\cite{LHT_Low,Hubisz_Meade,Belyaev_et_al,Choudhury,Hubisz_et_al}.

The masses of the heavy gauge bosons in the LHT are given by 
\be \label{gauge_boson_masses}
M_{W_H} = M_{Z_H} = g f \left(1 - {v^2 \over 8 f^2} \right) \approx 0.65 f,
\qquad  
M_{A_H} = {f g^\prime \over \sqrt{5}} \left(1 - {5 v^2 \over 8 f^2} \right)  
\approx 0.16 f, 
\ee
where corrections ${\cal O}(v^2/f^2)$ are neglected in the approximate
numerical values. Thus these particles have masses of several hundreds
of GeV for $f \sim 1\tev$, although $A_H$, the heavy partner of the
photon, can be quite light, because of the small prefactor, and is
usually assumed to be the LTP. When T-parity is unbroken, the LTP is
stable, and gives rise to large MET signatures together with energetic
jets and leptons from the cascades of heavy T-odd particles at the
LHC.  The masses of the heavy, T-odd fermions are determined by
general $3\times 3$ mass matrices in the (mirror) flavor space,
$m_{q_H,l_H}^{ij} \sim \kappa_{q,l}^{ij} f$ with $i,j=1,2,3$. We
simplify our analysis by assuming that $\kappa_q^{ij} = \kappa_q
\delta^{ij}$. The parameter $\kappa_q \sim {\cal O}(1)$ thus
determines the masses of the heavy quarks in the following way:
\be \label{mirror_quark_masses}
m_{u_H} = \sqrt{2} \kappa_q f \left(1 - {v^2 \over 8 f^2} \right), 
\qquad 
m_{d_H} = \sqrt{2} \kappa_q f, 
\ee
thereby allowing the new heavy quarks to have masses ranging from
several hundreds of GeV to a TeV, for $f \sim 1\tev$.  Similarly, the
masses of the heavy leptons in the spectrum are determined by a common
parameter $\kappa_l$. For our study we have kept $\kappa_l=1$. Note
that these heavy quarks and leptons cannot be decoupled from the model
as there is an upper bound $\kappa \leq 4.8$ (for $f = 1\tev$)
obtained from 4-fermion operators~\cite{Hubisz_et_al}.

The $A_H$, however, can decay if T-parity is violated. This can happen
via the so-called Wess-Zumino-Witten (WZW) term~\cite{WZW}, which,
according to Ref.~\cite{Hill_Hill}, can be written as follows:
\be \label{WZW_term}
\Gamma_{\rm WZW} = {N \over 48 \pi^2} \left( \Gamma_0[\Sigma] + \Gamma[\Sigma,
A_l, A_r] \right). 
\ee
The functional $\Gamma_0[\Sigma]$ is the ungauged WZW term which
depends only on the non-linear sigma model field $\Sigma$. The term
$\Gamma[\Sigma, A_l, A_r]$ is the gauged part of the WZW term.The
explicit expressions for the functionals and the relation of the
fields $A_{l,r}$ to the gauge fields in the LHT can be found in
Ref.~\cite{Hill_Hill}. While these functionals are uniquely given by
the symmetry breaking pattern $SU(5) \to SO(5)$ and the gauged
subgroups in the LHT, the integer $N$ in Eq.~(\ref{WZW_term}) depends
on the ultraviolet (UV) completion of the LHT. In strongly coupled
underlying theories it will be related to the representation of the
fermions whose condensate acts as order parameter of the spontaneous
symmetry breaking. In the simplest case, $N$ will simply be the number
of `colors' in that UV completion, as is the case for the WZW term in
ordinary QCD. The overall coefficient $N / 48 \pi^2$ encapsulates the
effect of the chiral anomaly, which is a one-loop effect in the
corresponding high-scale theory.

As noted in Ref.~\cite{Hill_Hill}, the WZW term in
Eq.~(\ref{WZW_term}) is not manifestly gauge invariant. Gauge
invariance is violated by terms with an odd number of T-odd gauge
bosons, such as one of the form $\epsilon_{\mu\nu\rho\sigma} V_H^\mu
V^\nu \partial^\rho V^\sigma$, where $V_H$ is a T-odd gauge boson and
$V$ denotes a SM gauge boson. Such anomalous terms need to be
cancelled~\cite{Hill_Hill}. After the cancellation, the leading T-odd
interactions appear only at order $1/f^2$. For instance, we get a
vertex with one T-odd gauge boson and two SM gauge bosons from
$\epsilon_{\mu\nu\rho\sigma} (H^\dagger H / f^2) V_H^\mu V^\nu
\partial^\rho V^\sigma$, after the Higgs doublet $H$ gets a vacuum
expectation value $v$.

To leading order in $1/f$, the part of the WZW term containing one
neutral T-odd gauge boson is given, in unitary gauge, by
\bea
\lefteqn{\Gamma_n  =  {N g^2 g^\prime \over 48 \pi^2 f^2} \int d^4x \, (v +
  h)^2 \, \epsilon_{\mu\nu\rho\sigma} \times \nonumber} \\
& & \left[ -(6/5) A_H^\mu \left( c_w^{-2} Z^\nu \partial^\rho Z^\sigma +
  W^{+\nu} D_A^\rho W^{-\sigma} + W^{-\nu} D_A^\rho W^{+\sigma} + i(3 g x_w +
  g^\prime s_w) W^{+\nu} W^{-\rho} Z^\sigma \right) \right. \nonumber \\ 
& & \left. + t_w^{-1} Z_H^\mu \left( 2 c_w^{-2} Z^\nu \partial^\rho Z^\sigma +
  W^{+\nu} D_A^\rho W^{-\sigma} + W^{-\nu} D_A^\rho W^{+\sigma} - 2i (2gc_w +
  g^\prime s_w) W^{+\nu} W^{-\rho} Z^\sigma \right) \right]. \nonumber \\
& & 
\label{WZW_A_H}
\eea
Here $h$ is the physical Higgs boson, $D_A^\mu W^{\pm\nu} =
(\partial^\mu \mp i e A^\mu)W^{\pm\nu}$ and $s_w, c_w$ and $t_w$
denote the sine, cosine and tangent of the weak mixing angle,
respectively.  All T-violating vertices with up to four legs have been
tabulated in Ref.~\cite{Freitas_Schwaller_Wyler} and implemented into
a model file for CalcHEP 2.5~\cite{CalcHEP, CalcHEP_T_violation}.

If $A_H$ is heavy, the vertices in Eq.~(\ref{WZW_A_H}) lead to its
decay into a pair of $Z$-bosons or into $W^+ W^-$ with a decay width
of the order of eV~\cite{Barger_Keung_Gao}. On the other hand, if
$M_{A_H} < 2 M_W$, the heavy photon cannot decay into on-shell SM
gauge bosons. It could still decay into (one or two) off-shell SM
gauge bosons, but for low masses loop-induced decays into SM fermions
will dominate. In fact, as discussed in
Ref.~\cite{Freitas_Schwaller_Wyler}, the T-violating vertices can
couple the $A_H$ to two SM fermions via a triangle loop. But since the
corresponding one-loop diagrams are logarithmically divergent, one
needs to add counterterms to the effective Lagrangian of the form
\bea
{\cal L}_{\rm ct} & = & \bar{f} \gamma_\mu \left( c_L^f P_L + c_R^f P_R
\right) f A_H^\mu, \label{counterterms} \\ 
c_i^f & = & c_{i,\epsilon}^f \left( {1 \over \epsilon} + \log(\mu^2) +
\order{1} \right), \label{coeff_CT}
\eea
where $P_{L,R} = (\mathbf{1} \mp \gamma_5)/2$.

The coefficients $c_i^f(\mu)$ of the counterterms can be estimated by
naive dimensional analysis or naturalness arguments.  The coefficients
$c_{i,\epsilon}^f$ are explicitly listed in
Ref.~\cite{Freitas_Schwaller_Wyler} and we have included the vertices
from Eq.~(\ref{counterterms}) in the CalcHEP model file.

The prefactor of the WZW term, $N/48\pi^2$, is of the size of a
one-loop effect, thus the coupling of $A_H$ and other T-odd gauge
bosons to SM fermions via a triangle-loop is effectively 2-loop
suppressed. Therefore these T-violating couplings will not affect the
production mechanism of T-odd particles and their cascade decays at
colliders, or EW precision
observables~\cite{Freitas_Schwaller_Wyler}. In particular, T-parity
violation should still satisfy the EW data with a rather small
scale~$f$.  It is only in decays of the $A_H$ that the anomaly term
acquires phenomenological importance.

Let us briefly note here how four or higher lepton events can arise in
LHT-TPV. All cascades initiated though the production of T-odd
particles (notably the heavy T-odd quarks that are produced via strong
interaction) lead to a pair of $A_H$. If kinematically allowed, the
principal decay modes of the $A_H$ are $A_H\rightarrow WW^{(*)}$ or
$A_H\rightarrow ZZ^{(*)}$. Thus, four leptons can come from the
leptonic decays of two Z-bosons, four W's, or two W's and one Z-boson
(every event will have two $A_H$'s produced at the end the
cascades). We shall also discuss about the other possibilities
section~\ref{subsec:invmass-2l}, where the loop-induced decay of each
$A_H$ to lepton pairs might also lead to four-lepton
events~\cite{Freitas_Schwaller_Wyler,Mukhopadhyay}.  One can also have
additional leptons from the cascades if the initially produced heavy
quarks can cascade decay to T-odd gauge bosons ($W_H^\pm$, $Z_H$) or
T-odd leptons ($l_H^\pm$, $\nu_H$). The two $A_H$'s can further give
rise to five lepton events if one of them decays to four-leptons (via
$ZZ$) and the other decays to $W^+W^-$, one of which then decays
semi-leptonically.

\section*{Acknowledgments} 

This work was partially supported by funding available from the
Department of Atomic Energy, Government of India, for the Regional
Centre for Accelerator-based Particle Physics (RECAPP), Harish-Chandra
Research Institute. SM and BM acknowledge the hospitality of the
Department of Theoretical Physics, Indian Association for the
Cultivation of Science, Kolkata.

\vspace{0.5cm}

\end{document}